\documentclass[prd,twocolumn,showpacs,showkeys,preprintnumbers,floatfix,
  nofootinbib,superscriptaddress]{revtex4-2}
\usepackage{amsfonts} % AMS
\usepackage{amssymb} % AMS
\usepackage{amsmath} % AMS
\usepackage{graphicx} % Include figure files
\usepackage{array} % array
\usepackage{dcolumn} % Align table columns on decimal point
\usepackage{bm} % bold math
\usepackage{latexsym} % latex symbols
\usepackage{longtable} % long tables
\usepackage{hyperref} % hypertext links 
\usepackage{multirow}
\usepackage{xcolor}
\usepackage{subfigure}
\usepackage{lipsum}
\usepackage{mathtools}
\usepackage{slashed}
\usepackage{comment}
\newcommand{\dslash}[1]{\text{$\not \!\! #1$}}

\newcommand{\GeV}{\,\mathop{\rm GeV}\nolimits}

\newcommand{\BtoDst}{\bar{B} \to D^\ast \ell \bar{\nu}}
\newcommand{\BtoD}{\bar{B} \to D \ell \bar{\nu}}
\newcommand{\BtoPi}{\bar{B} \to \pi \ell \bar{\nu}}
\newcommand{\BstoK}{\bar{B}_s \to K \ell \bar{\nu}}

\newcommand{\Vcb}{|V_{cb}|}
\newcommand{\Vub}{|V_{ub}|}

\newcommand{\epsK}{\varepsilon_{K}}

\newcommand{\com}[1]{} % delete it

\allowdisplaybreaks

\begin{document}

\title{ Improvement of heavy-heavy and heavy-light currents with the
  Oktay-Kronfeld action }
\author{Jon A. Bailey}
\affiliation{
  ISED, UIC, Yonsei University,
  Incheon 21983, South Korea
}
\author{Yong-Chull Jang}
%
%\email[E-mail: ]{yj2649@columbia.edu}
%
\affiliation{
Columbia University
Department of Physics
538 West 120th Street
New York, NY 10027, USA }
\author{Sunkyu Lee}
%
%%\email[E-mail: ]{wlee@snu.ac.kr}
%
% \homepage[Home page at ]{http://lgt.snu.ac.kr/}
%
%
\affiliation{
  Lattice Gauge Theory Research Center, FPRD, and CTP, 
  Department of Physics and Astronomy,
  Seoul National University, Seoul 08826, South Korea
}
\author{Weonjong Lee}
\email[E-mail: ]{wlee@snu.ac.kr}
%
% \homepage[Home page at ]{http://lgt.snu.ac.kr/}
%
%
\affiliation{
  Lattice Gauge Theory Research Center, FPRD, and CTP, 
  Department of Physics and Astronomy,
  Seoul National University, Seoul 08826, South Korea
}
\author{Jaehoon Leem}
\email[E-mail: ]{leemjaehoon@kias.re.kr}
\affiliation{
  School of Physics, Korea Institute for Advanced Study (KIAS),
  Seoul 02455, South Korea
}  
\collaboration{LANL-SWME Collaboration}
\date{\today}
\begin{abstract}
  The CKM matrix elements $\Vcb$ and $\Vub$ can be obtained by
  combining data from the experiments with lattice QCD results for the
  semi-leptonic form factors for the $\BtoDst$ and $\BtoPi$ decays.
  It is highly desirable to use the Oktay-Kronfeld (OK) action for the
  form factor calculation on the lattice, since the OK action is
  designed to reduce the heavy quark discretization error down to the
  $\mathcal{O}(\lambda^4)$ level in the power counting rules of the
  heavy quark effective theory (HQET).
  Here, we present a matching calculation to improve heavy-heavy and
  heavy-light currents up to the $\lambda^3$ order in HQET, the same
  level of improvement as the OK action.
  Our final results for the improved currents are being used in a lattice
  QCD calculation of the semi-leptonic form factors for the $\BtoDst$
  and $\BtoD$ decays.
\end{abstract}
\maketitle
%
%
%

%------------
% SECTION 1.
%\input{intro}
%------------
%
\section{Introduction}
\label{sec:intr}

The Cabibbo-Kobayashi-Maskawa (CKM) matrix contains four of the
fundamental parameters of the Standard Model (SM) which describes
flavor-changing phenomena and CP violation \cite{ Buchalla:1995vs,
  Winstein:1992sx}.

The CKM matrix is a $3\times 3$ unitary matrix, and $\Vcb$ is a CKM
matrix element which describes flavor-changing weak interactions
between bottom and charm quarks.
$\Vcb$ is an important quantity in particle physics.
It constrains one side of the unitarity triangle through the ratio
$\Vub/\Vcb$.
It gives the dominant uncertainty in the determination of the CP
violation parameter $\epsK$ in the neutral kaon system, where there is
currently tension between the SM and experiment \cite{
  Bailey:2018feb}.

There are two competing and independent methods to determine $\Vcb$:
one is to derive $\Vcb$ from the exclusive decays ($\BtoDst$ and
$\BtoD$) and the other is to obtain $\Vcb$ from the inclusive decays
($B \to X_c \ell \nu$).
There exists currently $3\sigma \sim 4\sigma$ tension between the
exclusive $\Vcb$ and the inclusive $\Vcb$ \cite{ HFLAV:2019otj,
  FermilabLattice:2021cdg}, which makes the study of $\Vcb$ even more
interesting.

Another motivation to study the exclusive decays ($\BtoDst$ and
$\BtoD$) is the tension in $R(D^{(\ast)})$ between the SM theory and
experiment \cite{ Amhis:2016xyh}.
An update from HFLAV \cite{ Amhis:2016xyh} gave the combined tension in
$R(D)$ and $R(D^\ast)$ to be about $3.8\sigma$.
A recent report from HFLAV \cite{ HFLAV:2019otj} and BELLE \cite{
  Thalmeier:2019mhn} claimed that the tension is about $3\sigma$.
Hence, more precise determination of the semi-leptonic form factors
for the exclusive decays will be important to confirm or dismiss a
potential new physics possibility.

When we determine $\Vcb$ from the exclusive decays such as $\BtoDst$,
there are two different sources of uncertainty: One comes from the
theory, and the other comes from experiment.
Basically the experiments determine $\Vcb \cdot |\mathcal{F}(1)|$ and
the theory determines the form factors $|\mathcal{F}(1)|$.
The dominant uncertainty in the calculation of the semi-leptonic form
factors $|\mathcal{F}(1)|$ comes from the heavy-quark discretization
\cite{ Bailey:2014tva}.
Hence, it is essential to reduce the heavy-quark discretization error
as much as possible in order to achieve higher precision in
$|\mathcal{F}(1)|$.

It is challenging to reduce the discretization errors for $b$ and $c$ quarks,
since the heavy quark masses are comparable with or greater than the inverse of the lattice spacing $1/a$.
The Symanzik improvement program~\cite{Symanzik:1979ph} does not work
for $a m_Q \approx 1$.
The Fermilab formalism \cite{ ElKhadra:1996mp} makes it possible to
control the discretization errors of bottom and charm quarks on 
relatively coarse lattices.
In the Fermilab formalism, the lattice artifacts for heavy quarks are
bounded in the limit of $m_Q a \to \infty$, and they can be reduced
systematically by tuning coefficients of the action.
With a non-relativistic interpretation of the Wilson action, one can
match the lattice theory to continuum QCD using the heavy-quark
effective theory (HQET) for heavy-light systems \cite{ Eichten:1989zv,
  Georgi:1990um, Grinstein:1990mj} or non-relativistic QCD (NRQCD) for
quarkonia \cite{ Caswell:1985ui, Lepage:1992tx}.
Here we can estimate the lattice artifacts due to neglecting the
truncated higher order terms by using the power counting of HQET or
NRQCD.

The Fermilab action includes the dimension five operators of the
Wilson clover action and is improved up to the $\lambda^1$ order in
HQET \cite{ ElKhadra:1996mp}.
The Oktay-Kronfeld (OK) action is an extension of the Fermilab action
and is improved up to the $\lambda^3$ order in HQET \cite{
  PhysRevD.78.014504}.
In order to calculate weak matrix elements while taking advantage of
the full merits of the OK action, it is essential to improve also the
flavor-changing currents up to the $\lambda^3$ order at the tree
level.
In this paper we explain additional operators needed to improve the
currents up to the $\lambda^3$ order and a matching calculation to
determine the coefficients for these operators.
The resulting improved currents can be used to calculate the semi-leptonic form
factors for the $\BtoDst$ and $\BtoD$ decays \cite{ Bhattacharya:2018gan,
Bhattacharya:2020xyb}.

In Section \ref{sec:ok} we briefly review the Fermilab formalism and
show the explicit forms of the Fermilab and OK actions.
In Section \ref{sec:current-improvement} we introduce an approach to
current improvement and build up the improved current.
In Section \ref{sec:matching-calculation} we explain the matching
calculations and determine the improvement parameters, the
coefficients for the improved current operators.
In Section \ref{sec:hqet-description} we present an interpretation of
the matching calculation based on HQET.
The HQET interpretation clarifies the structure of the matching
conditions and provides a cross-check.
In Section \ref{sec:results} we present the results for the
improvement parameters and discuss their continuum and static
limits.
In Section \ref{sec:conclusion} we conclude.
The appendices contain technical details on the matching calculations
and comparison of the continuum limit with results from the
Symanzik program.

Preliminary results for the improved currents were presented in
\cite{Bailey:2017zgt}.

%%%\wlee{EDIT by wlee}

%
%-------------
% SECTION 2.
%\input{ok}
%-------------
%
\section{Lattice Actions for Heavy Quarks}
\label{sec:ok}

The Fermilab method \cite{ElKhadra:1996mp} is used to
systematically improve lattice gauge theories with Wilson
quarks~\cite{Wilson:1975id} with masses comparable to the lattice
cutoff, $a m_Q \simeq 1$.
Symanzik's original local effective description of lattice gauge
theory \cite{ Symanzik:1979ph} assumes $a m_Q \ll 1$, and so it does
not apply to heavy quarks.
Instead, HQET and NRQCD can be used as alternative effective
field theories to describe the lattice artifacts of heavy quarks
\cite{ Kronfeld:2000ck, Harada:2001fi, Harada:2001fj}.
A dual expansion in $\lambda \sim \Lambda/(2 m_Q) \sim
a\Lambda$ is used to construct the $\mathcal{O}(\lambda^1)$ action of
effective-continuum HQET.
Using a generalized version of Symanzik's effective field
  theory together with effective-continuum HQET and NRQCD, an improved
  version of the Fermilab action was developed in Ref.~\cite{
    PhysRevD.78.014504}.  
It is called the OK action, which includes improvement terms
  through $\mathcal{O}(\lambda^3)$.  

The Fermilab method begins with the observation that time-space
axis-interchange symmetry need not be respected to tune the lattice
action and currents to the renormalized trajectory~\cite{ Wilson:1973jj}.
For systems with heavy quarks, Ref.~\cite{
  ElKhadra:1996mp} introduced independent, mass-dependent couplings
for the spatial and temporal parts of the clover term~\cite{
  Sheikholeslami:1985ij} and pointed out the sufficiency of including
only spatial terms at higher order, without altering the Wilson time
derivative.
Constructing the transfer matrix and deriving the Hamiltonian, it is
shown that the discretization errors remain bounded as 
$a m_Q \to \infty$.

The analysis of the lattice Hamiltonian also led to introducing an improved
quark field for flavor-changing currents \cite{ElKhadra:1996mp}.
Constructing flavor-changing currents with the improved quark fields,
the coefficients of the improvement terms can be determined uniquely by matching
two-quark matrix elements.
In Refs.~\cite{Harada:2001fi,Harada:2001fj}, it was proven that for improvement
through $\mathcal{O}(\lambda)$ in HQET it is sufficient to match the improved
field at tree-level.

The equivalence of the lattice theory and HQET can be expressed by the
relation
\begin{align}
S_\text{lat} \doteq 
S_{\text{HQET}} =\int d^4 x ~ \mathcal{L}_{\text{HQET}}\,,
\label{eq:LEAFF}
\end{align}
where the symbol $\doteq$ means that, in the regime where both
theories hold, all physical amplitudes with external states on shell
are equal to each other, and
\begin{align}
\mathcal{L}_{\text{HQET}} = \bar{h}^{+}\Big( D_4 +m_1 
- \frac{\boldsymbol{D}^2}{2m_2}
+\frac{z_B i\boldsymbol{\sigma \cdot B}}{2m_B}\Big) h^{+} +\cdots\,, 
\label{eq:sec.II-fermilab-hqet}
\end{align}
where $z_B$ is the matching coefficient for the chromomagnetic term, and
$m_1$, $m_2$, and $m_B$ are the rest, kinetic, and 
chromomagnetic masses of the quark, respectively.
Here, $h^{+}$ is a heavy-quark field which satisfies $\gamma_4 h^{+}
= h^{+}$.
When we consider matching between the lattice theory and HQET, the 
rest mass $m_1$ makes no difference because it does not affect the 
energy splittings and the matrix elements \cite{Kronfeld:2000ck}.
The bare mass (or the hopping parameter) is determined by demanding
that the kinetic mass $m_2$ be equal to the physical mass.

The explicit formula of the Fermilab action \cite{ElKhadra:1996mp} is
\begin{align}
S_\text{Fermilab} = S_0 +S_B +S_E,
\end{align}
where 
\begin{align}
S_0  = 
&a^4 \sum_x \big[m_0 \bar{\psi}(x)\psi(x) 
+ \bar{\psi}(x) \gamma_4D_{\text{lat},4} \psi(x) 
\nonumber \\
&+\zeta 
\bar{\psi}(x)\boldsymbol{ \gamma }\cdot \boldsymbol{D}_{\text{lat}} \psi(x)
-\frac{1}{2}a \bar{\psi}(x) \Delta_4 \psi (x) 
\nonumber \\
&-\frac{1}{2}r_s \zeta a  \bar{\psi}(x)  \Delta^{(3)} \psi (x) \big]\,,
\label{eq:fermi-act-1}
\end{align}
where $m_0$ is a bare quark mass, the parameter $\zeta$ breaks
axis-interchange symmetry if $\zeta\neq 1$, and $r_s$ is the Wilson
parameter for the spatial directions.
The lattice covariant derivative operators are
\begin{align}
D_{\text{lat},\mu} \psi   
&= (2a)^{-1}(T_\mu -T_{-\mu})\psi\,,  
\\
\Delta_\mu \psi   
&=  a^{-2}(T_\mu +T_{-\mu}-2)\psi\,, 
\\
\Delta^{(3)} \psi   
&=  \sum_{i=1}^3 \Delta_i \psi\,,
\end{align}
where the covariant translation is defined by
\begin{align}
  T_{\pm\mu} \psi(x) &= U_{\pm\mu}(x)\psi(x\pm a\hat{\mu})\,,
  \\
  U_{\pm\mu} (x) &= U(x, x\pm a\hat{\mu})\,,
\end{align}
where $\pm \mu$ represents the positive and negative directions along
the $\mu$-axis, and $\hat{\mu}$ is a unit vector along the $\mu$-axis.
The dimension five operators $S_B$ and $S_E$ are
\begin{align}
S_B & = -\frac{1}{2}c_B \zeta a^5 \sum_x \bar{\psi}(x) i 
\boldsymbol{\Sigma} \cdot \boldsymbol{B}_{\text{lat}}\psi(x), 
\\
S_E & = -\frac{1}{2}c_E \zeta a^5 \sum_x \bar{\psi}(x) 
\boldsymbol{\alpha} \cdot \boldsymbol{E}_{\text{lat}}\psi(x) \,.
\end{align}
Here the chromomagnetic and the chromoelectric fields are
\begin{align}
B_{\text{lat},i} 
&= \frac{1}{2}\epsilon_{ijk}F^{\text{lat}}_{jk}, 
\qquad 
E_{\text{lat},i} 
= F^{\text{lat}}_{4i},
\end{align}
with the clover field-strength tensor
\begin{align}
F^\text{lat}_{\mu \nu} 
= \frac{1}{8a^2} \sum_{\bar{\mu}=\pm \mu , \atop\bar{\nu} = \pm \nu}
\text{sign}(\bar{\mu})\text{sign}(\bar{\nu})T_{\bar{\mu}}T_{\bar{\nu}}T_{-\bar{\mu}}T_{-\bar{\nu}}
-\text{h.c.}\,.
\end{align}
Here $\text{sign}(\bar{\mu})=\pm 1 $ for $\bar{\mu}=\pm \mu$.

The OK action \cite{PhysRevD.78.014504} includes counter-terms up to 
$\lambda^3$ order, incorporating all dimension six and some dimension 
seven bilinear operators.
The OK action is
\begin{align}
S_{\text{OK}} = S_0 + S_B + S_E + S_6 + S_7\,,
\end{align}
where $S_6$ ($S_7$) represents counter-terms of dimension
six (seven). Explicitly,
\begin{align}
\label{eq:dimension-six-OK}
&S_6 
=a^6\sum_x \bar{\psi}(x)\Big[
c_1 
\sum_i \gamma_i D_{\text{lat},i}\Delta_{\text{lat},i}
+ c_2 \{  \boldsymbol{\gamma} \cdot \boldsymbol{D}_{\text{lat}}, 
\Delta^{(3)}\}  
\nonumber \\
&+ c_3 \{ \boldsymbol{\gamma} \cdot \boldsymbol{D}_{\text{lat}},
i \boldsymbol{\Sigma} \cdot \boldsymbol{B}_{\text{lat}}\}
+ c_{EE} \{\gamma_4 D_{\text{lat},4},
\boldsymbol{\alpha} \cdot \boldsymbol{E}_{\text{lat}}\}
\Big]
\psi(x)
\,,
\end{align}
and
\begin{align}
\label{eq:dimension-seven-OK}
&S_7  
= a^7 \sum_x \bar{\psi}(x) \sum_i\Big[ c_4 \Delta_{i}^2 
+ c_5 \sum_{j \ne i } \{i\Sigma_i B_{\text{lat},i}, \Delta_{j}\}\Big]\psi(x)
\,.
\end{align}
The coefficients $\{c_i\}$ are determined by matching the dispersion 
relation, interaction with a background field, and Compton scattering 
amplitude at tree level.

Taking redundant operators into account, the operators in
Eqs.~\eqref{eq:dimension-six-OK} and~\eqref{eq:dimension-seven-OK} are
a complete set for matching through $\mathcal{O}(\lambda^3)$ at tree
level.
In general, at dimension six, there are contributions from not only 
bilinears, but also four-quark operators such as
\begin{align}
  &[\bar{Q}\Gamma Q][\bar{Q}\Gamma Q]\,,
  \label{eq:hh}
  \\
  &[\bar{Q}\Gamma Q] \displaystyle \sum_{f} [\bar{q}_f\Gamma q_f] \,,
  \label{eq:hl}
\end{align}
where $Q$ represents heavy quarks, and $q_f$ represents light quarks
with flavor $f$.
In the heavy-light system, however, four-quark operators of the type
in Eq.~\eqref{eq:hh} contribute to physical matrix elements only
through heavy-quark loops, and so contributions from these operators
are suppressed by at least an additional factor of $\lambda^2$
\cite{PhysRevD.78.014504}; such operators are omitted from the OK
action.
When [heavy quark]-[light quark] scattering is matched at tree level,
one finds that the tree-level coupling of four-quark operators of the
type in Eq.~\eqref{eq:hl} is proportional to a redundant coupling of
the pure-gauge action, and can be eliminated by adjusting this
coupling \cite{PhysRevD.78.014504}.
Thus, the four-quark operators are neglected, and the OK action has
only six new bilinear operators.
%

%-----------------
% SECTION 3.
%\input{curr_impr}
%-----------------
%
\section{
Improvement terms for the lattice heavy quark currents
}
\label{sec:current-improvement}

In the calculation of hadronic matrix elements for $\bar{B} \to
D^{(*)} \ell \bar{\nu}$ decay, heavy-quark discretization errors
come from both the hadronic states and the flavor-changing
currents \cite{ Kronfeld:2000ck, Harada:2001fj}.
Using the OK action for $b$ and $c$ quarks, we expect the hadronic
states of the $B$ and $D^{(*)}$ mesons to be improved up to
$\lambda^3$ order by the action itself.
To take full advantage of the OK action for $b$ and $c$ quarks, we
must improve the flavor-changing currents up to $\lambda^3$ order, the
level of improvement of the OK action.
Here we explain how to improve the currents up to $\lambda^3$ order
using HQET.

The current improvement to first order in $\lambda$
was studied in \cite{ElKhadra:1996mp, Kronfeld:2000ck, Harada:2001fj}.
If one neglects loop corrections, the current improvement can be done by 
introducing an improved quark field \cite{ElKhadra:1996mp,  Harada:2001fj}.
\begin{align}
\label{eq:improved-current}
V^{\text{lat}}_\mu &= \bar{\Psi}_{Ic}\gamma_\mu \Psi_{Ib}, 
\\
A^{\text{lat}}_\mu &= \bar{\Psi}_{Ic}\gamma_\mu \gamma_5 \Psi_{Ib}, 
\label{eq:improved-currents-by-improved-field}
\end{align}
where $\Psi_{If}$ is ($f=b,c$)
\begin{align}
\label{eq:improved-quark-field-lambda}
\Psi_{If}(x) 
\equiv e^{m_{1f}a/2} [ 1+ ad_{1f} \boldsymbol{\gamma \cdot D_{\text{lat}}}]\psi_f(x) \,.
\end{align}
Here, the normalization factor $e^{m_{1f}a/2}$ is introduced to
cancel out the field renormalization of the lattice quark fields :
$m_{1f}a = \log(1+m_{0f}a)$ is the rest mass at tree level ($f=b,c$).
The parameter $d_1$ is an improvement parameter to be determined by a
matching condition.
In \cite{ElKhadra:1996mp, Harada:2001fj}, it is shown that 
introducing the improved quark field Eq.~\eqref{eq:improved-quark-field-lambda}
is enough for the current improvement at tree level.

Here we would like to extend the idea of the improved quark field
to $\mathcal{O}(\lambda^3)$.
We need to find a complete set of operators up to dimension six.
The continuum Foldy-Wouthuysen-Tani (FWT) transformation \cite{
  Foldy:1949wa,doi:10.1143/ptp/6.3.267} is a good starting point.

Let us review how to derive the HQET Lagrangian from
the QCD Lagrangian.
The fermionic part of the QCD Lagrangian in Euclidean space is

\begin{align}
  &\mathcal{L}_{\text{Dirac}} = -\bar{Q}(\dslash{D}+m)Q \,,
  \label{eq:QCD-L}
\end{align}
where $Q$ is a heavy quark field with mass $m$.
At tree level, the HQET Lagrangian can be derived by using a
  FWT transformation, which decouples quark and anti-quark.
The FWT transformation up to $1/m^4$ order is
\begin{align}
\label{eq:HQET-FWT-1}
&Q =
\bigg[ 1- \frac{1}{2m} \boldsymbol{\gamma \cdot D}
+\frac{1}{8m^2}(\boldsymbol{\gamma \cdot D})^2
+\frac{1}{4m^2}\boldsymbol{\alpha}\cdot \boldsymbol{E}
\nonumber \\
&
-\frac{3(\boldsymbol{\gamma \cdot D})^3}{16m^3}
-\frac{\boldsymbol{\gamma}\cdot \boldsymbol{D}\boldsymbol{\alpha}\cdot \boldsymbol{E}}{8m^3}
-\frac{\{\gamma_4D_4,\boldsymbol{\alpha}\cdot \boldsymbol{E}\}}{8m^3}
\nonumber \\
&
+\frac{11(\boldsymbol{\gamma \cdot D})^4}{128m^4}
+\frac{3(\boldsymbol{\gamma \cdot D})^3 \gamma_4 D_4}{16m^4}
\nonumber \\
&
+\frac{(\boldsymbol{\gamma \cdot D})^2 \gamma_4 D_4 (\boldsymbol{\gamma \cdot D})}{8m^4}
+\frac{3(\boldsymbol{\gamma \cdot D}) \gamma_4 D_4 (\boldsymbol{\gamma \cdot D})^2}{32m^4}
\nonumber \\
&
+\frac{5\gamma_4 D_4 (\boldsymbol{\gamma \cdot D})^3}{32m^4}
+\frac{\boldsymbol{\gamma \cdot D}\{\gamma_4 D_4 , \boldsymbol{\alpha \cdot E}\}}{16m^4}
+\frac{(\boldsymbol{\alpha \cdot E})^2}{32m^4}
\nonumber \\
&+\frac{1}{16m^4}\big\{ \gamma_4 D_4 , \{ \gamma_4 D_4, \boldsymbol{\alpha \cdot E} \} \big\}
\bigg]h+\mathcal{O}(1/m^5)\,.
\end{align}
The corresponding HQET Lagrangian up to $1/m^3$ order is
\begin{align}
\mathcal{L}_{HQ}&=\bar{h}^+ \bigg[ -D_4 -m
+\frac{1}{2m}\boldsymbol{D}^2
+\frac{i}{2m}\boldsymbol{\sigma \cdot B}
\nonumber \\
&+\frac{  \boldsymbol{D}\cdot\boldsymbol{E}
-\boldsymbol{E}\cdot\boldsymbol{D} }{8m^2}
+\frac{i \boldsymbol{\sigma} \cdot
\big(\boldsymbol{D}\times \boldsymbol{E}-\boldsymbol{E}\times \boldsymbol{D}\big) }{8m^2}
\nonumber \\
&+\frac{1}{8m^3} (\boldsymbol{\sigma}\cdot\boldsymbol{D})^4
-\frac{1}{8m^3} (\boldsymbol{\sigma \cdot E})^2
\bigg]h^+ + \ldots,
\label{eq:HQET-lambda-third}
\end{align}
where $h$ is the heavy quark field in the rest frame of the heavy
quark, with quark field $h^+$ and anti-quark field $h^-$:
\begin{align}
  h^\pm = \frac{1\pm \gamma_4}{2}h \,.
\end{align}
In Eq.~\eqref{eq:HQET-lambda-third}, we drop terms with the
  anti-quark field $h^-$ for simplicity.
Eq.~\eqref{eq:HQET-lambda-third} is consistent with the NRQCD
Lagrangian at the tree-level \cite{ Manohar:1997qy}.
A study on extending Eq.~\eqref{eq:HQET-FWT-1} to arbitrary higher order is
given in Ref.~\cite{Balk:1993ev}.

Taking the continuum FWT transformation as an ansatz, we
  introduce the $\mathcal{O}(\lambda^3)$-improved quark field on the
  lattice as follows,
\begin{align}
\label{eq:improved-current-lambda-third}
\Psi_{I}&(x) 
= e^{m_{1}a/2}
\Big[1+ad_{1} \boldsymbol{\gamma} \cdot \boldsymbol{D}_{\text{lat}}  
+\frac{1}{2}a^2d_2 \Delta^{(3)} 
\nonumber \\
&
+\frac{1}{2}a^2 d_B i\boldsymbol{\Sigma}\cdot\boldsymbol{B}_{\text{lat}}
+\frac{1}{2}a^2 d_E \boldsymbol{\alpha} \cdot \boldsymbol{E}_{\text{lat}}
\nonumber \\
&
+ a^3 d_{EE}\{\gamma_4 D_{4\text{lat}}, \boldsymbol{ \alpha} \cdot \boldsymbol{E}_{\text{lat}}\} 
+ \frac{1}{6}a^3 d_{3} \gamma_i D_{\text{lat},i} \Delta_{i} 
\nonumber \\
&
+ \frac{1}{2}a^3 d_{4}\{\boldsymbol{ \gamma} \cdot \boldsymbol{D}_{\text{lat}},\Delta^{(3)} \}
+ a^3 d_{5}\{\boldsymbol{ \gamma} \cdot \boldsymbol{D}_{\text{lat}} ,
i \boldsymbol{ \Sigma} \cdot \boldsymbol{B}_{\text{lat}}\}
\nonumber \\
&
+a^3 d_{r_E} \{ \boldsymbol{\gamma} \cdot \boldsymbol{D}_{\text{lat}}, 
\boldsymbol{\alpha} \cdot \boldsymbol{E}_{\text{lat}}\} 
+a^3 d_6 [\gamma_4 D_{4\text{lat}} ,\Delta^{(3)} ]
\nonumber \\
&
+a^3 d_7 [\gamma_4 D_{4\text{lat}}, i\boldsymbol{\Sigma}\cdot \boldsymbol{B}_{\text{lat}} ]
\Big] \psi(x).
\end{align}
Here note that the terms up to dimension five are identical to
  those introduced in Ref.~\cite{ElKhadra:1996mp}.
To compare Eq.~\eqref{eq:improved-current-lambda-third} with the 
continuum FWT transformation in Eq.~\eqref{eq:HQET-FWT-1}, let us
rearrange terms up to $\mathcal{O}(1/m^3)$ in Eq.~\eqref{eq:HQET-FWT-1}
as follows.
\begin{align}
\label{eq:HQET-FWT-2}
Q &=
\Big[ 1- \frac{1}{2m} \boldsymbol{\gamma \cdot D} 
+\frac{1}{8m^2}\boldsymbol{D}^2
+\frac{i}{8m^2}\boldsymbol{\Sigma}\cdot \boldsymbol{B}
+\frac{1}{4m^2}\boldsymbol{\alpha}\cdot \boldsymbol{E}
\nonumber \\
&-\frac{\{\gamma_4D_4,\boldsymbol{\alpha}\cdot \boldsymbol{E}\}}{8m^3}
-\frac{3\{\boldsymbol{\gamma \cdot D},\boldsymbol{D}^2\}}{32m^3}
-\frac{3\{\boldsymbol{\gamma \cdot D},i\boldsymbol{\Sigma}\cdot \boldsymbol{B}\}}{32m^3}
\nonumber \\
&-\frac{\{\boldsymbol{\gamma}\cdot \boldsymbol{D}, 
\boldsymbol{\alpha}\cdot \boldsymbol{E}\}}{16m^3}
+\frac{[\gamma_4 D_4,\boldsymbol{D}^2]}{16m^3}
+\frac{[\gamma_4 D_4,i\boldsymbol{\Sigma} \cdot \boldsymbol{B}]}{16m^3}
\Big]h
\nonumber \\
&=  \mathbf{U}_b \cdot h_b \,.
\end{align}
%
%We note that the $d_3$ term is absent in Eq.~\eqref{eq:HQET-FWT-2}.
%
All the terms in Eq.~\eqref{eq:improved-current-lambda-third} except
the $d_3$ terms have corresponding terms in Eq.~\eqref{eq:HQET-FWT-2}.
The $d_3$ term is necessary to remove rotational symmetry breaking
effects on the lattice.
%
%

%-----------------
% SECTION 4.
%\input{matching}
%-----------------
%
\section{Matching Calculation}
\label{sec:matching-calculation}

Now, we need to determine the improvement parameters $d_i$ in
Eq.~\eqref{eq:improved-current-lambda-third}.
%
%In the Sec.\ref{sec:matching-calculation}, we will introduce a 
%matching condition and perform a matching calculation.
%
There are many relevant matrix elements for matching.
If we choose the simplest two-quark matrix element,
$\langle c(p^\prime , s^\prime)|
J_\mu|b(p,s)\rangle$ (with $J = V, A$), we can 
determine $d_1$--$d_4$, but cannot determine the rest.
To determine the remaining parameters, we match 
matrix elements with one-gluon exchange.
We can choose the four-quark matrix element
$\langle \ell(p_2,s_2) c(p^\prime,s^\prime)| \;
J_\mu \; | b(p,s) \ell(p_1,s_1)\rangle \,, $
with one spectator light quark $\ell$ which exchanges a gluon with
heavy quarks.
%In the next section, we explain details of the matching procedure.
%
In the following two subsections, we show matching calculations with
two-quark and four-quark matrix elements, respectively.

\subsection{Matching two-quark matrix element}
\label{subsec:matching-two-quark}

Let us consider the following matrix element of lattice and continuum 
QCD
\begin{align}
  \langle c(p^\prime,&s^\prime)| \;
  \bar{\Psi}_{Ic} \Gamma \Psi_{Ib}
  \; | b(p,s) \rangle_\text{lat}
\nonumber \\
&  =
%%%  \longleftrightarrow
  \langle c(p^\prime,s^\prime)| \;
  \bar{c} \Gamma b
  \; | b(p,s) \rangle_\text{con} \,,
  \label{eq:2q-ME}
\end{align}
where $\Gamma = \gamma_\mu, \gamma_\mu \gamma_5$ represents the Dirac matrices
of the flavor-changing currents, and $\Psi_{Ib}$ and $\bar{\Psi}_{Ic}$ are the
improved quark fields defined in Eq.~\eqref{eq:improved-current-lambda-third}.
In the equations of this and the following sections, we set $a=1$
for notational convenience.

At tree level, the difference between lattice and continuum 
matrix elements comes from the spinors and normalization factors.
\begin{align}
\sqrt{\frac{m}{E}}u(p,s) = \Big[1 -\frac{i\boldsymbol{\gamma \cdot p}}{2m}
&-\frac{\boldsymbol{p}^2}{8m^2}
+ \frac{3i(\boldsymbol{\gamma \cdot p})\boldsymbol{p}^2}{16m^3}\Big] 
\nonumber \\
&
\times u(0,s)
+\mathcal{O}(\boldsymbol{p}^4),
\label{eq:conti-spinor-expansion}
\end{align}
The corresponding spinor on the lattice can be expanded as follows 
\begin{align}
&\mathcal{N}(p)u^{\text{lat}}(p,s) = e^{-m_1/2}
\Big[1- \frac{i\zeta \boldsymbol{\gamma \cdot p}}{2\sinh m_1} 
-\frac{\boldsymbol{p}^2}{8m_X^2}
\nonumber \\
&
+ \frac{i}{6}\frac{3c_1+\zeta/2}{\sinh m_1}
\sum^3_{k=1}\gamma_k p^3_k 
+ \frac{3i(\boldsymbol{\gamma \cdot p})\boldsymbol{p}^2}{16m_Y^3} 
\Big] u(0,s)+\mathcal{O}(\boldsymbol{p}^4),
\label{eq:lattice-spinor-expansion}
\end{align}
where 
%$m_X$, $m_Y$, and $w_3$ are defined in Ref.~\cite{Bailey:2014jga} as
%
\begin{align}
% w_3 &\equiv \frac{3c_1+\zeta/2}{\sinh m_1}, 
% \\
  \frac{1}{8m_X^2} &\equiv \frac{\zeta^2}{8\sinh^2 m_1}
  + \frac{r_s \zeta}{4e^{m_1}},
 \\
 \frac{3}{16m_Y^3} &\equiv \frac{1}{2\sinh m_1}
 \Big\{2c_2 + \frac{1}{4} e^{-m_1}
 \Big[\zeta^2 r_s (2\coth m_1 +1) 
   \nonumber \\
   &+\frac{\zeta^3}{\sinh m_1}\Big(\frac{e^{-m_1}}{2\sinh m_1} -1 \Big)\Big]
 +\frac{\zeta^3}{4\sinh^2 m_1}
 \Big\}.
\end{align}
Here $\mathcal{N}(p)$ is the normalization factor for a spinor of
the external quark line on the lattice, while
$\sqrt{\dfrac{m}{E}}$ is that in the continuum.
Explicit formulas for $\mathcal{N}(p)$, $u^{\text{lat}}(p,s)$,
$u(p,s)$ are given in Appendix \ref{app:LFR-1}.

The matching condition can be expressed as
\begin{align}
\mathcal{N}_b(p)R^{(0)}_b(p) 
u^{\text{lat}}_b(p,s)  &= 
\sqrt{\frac{m_b}{E_b}}u_b(p,s),
\label{eq:two-match-subdiagram-1}
\\
\mathcal{N}_c(p^\prime)
\bar{u}^{\text{lat}}_c(p^\prime,s^\prime)
\bar{R}^{(0)}_c(p^\prime)  &= 
\sqrt{\frac{m_c}{E_c}}\bar{u}_c(p^\prime,s^\prime),
\label{eq:two-match-subdiagram-2}
\end{align}
where subscripts $b,c$ are introduced to 
distinguish bottom and charm. 
$R^{(0)}(p)$ represents the zero-gluon vertex, which contains 
kinetic corrections and the normalization factor from the 
improved quark field.
The explicit formula of $R^{(0)}(p)$ is given in Appendix \ref{app:LFR-1}.
The overall factor $e^{m_1/2}$ from the improved quark field (in
Eq.~\eqref{eq:improved-current-lambda-third}) cancels out the overall
factor $e^{-m_1/2}$ in Eq.~\eqref{eq:lattice-spinor-expansion}, which
leads to the matching condition of
Eq.~\eqref{eq:two-match-subdiagram-1}.  

Expanding in $\boldsymbol{p}a$ and comparing terms up to 
$\mathcal{O}(\boldsymbol{p}^3)$, one can determine $d_1, d_2, d_3$, and $d_4$.
For example, from matching in $\mathcal{O}(\boldsymbol{p})$ 
\cite{ElKhadra:1996mp,Harada:2001fj},
\begin{align}
d_1 & = \frac{\zeta}{2\sinh m_1}
-\frac{1}{2m}
=\frac{\zeta(1+m_0)}{m_0(2+m_0)}-\frac{1}{2m}.
\label{eq:d1-match}
\end{align}
The results for $d_2,d_3$, and $d_4$ are given in Sec.\ref{sec:results}.
Especially, the rotational symmetry breaking term with $d_3$ in 
Eq.~\eqref{eq:improved-current-lambda-third} eliminates the unwanted symmetry 
breaking term $\sum_{k=1}^3 \gamma_k p_k^3$ in 
Eq.~\eqref{eq:lattice-spinor-expansion}.

In tree level matching, the other improvement parameters do 
not contribute to the two-quark matrix element. 
One should choose matrix elements with external gluons or gluon exchange.
In the next subsection, we introduce a four-quark matrix element with
additional light spectator quarks, which includes a gluon exchange.  

\subsection{Matching four-quark matrix element}
\label{subsec:matching-four-quark}

Let us consider the following four-quark matrix element for matching:
\begin{align}
  \langle \ell(p_2,&s_2)
c(p^\prime,s^\prime)| \;
  \bar{\Psi}_{Ic} \Gamma \Psi_{Ib}
  \; | b(p,s) \ell(p_1,s_1)\rangle_\text{lat}
\nonumber \\
&  =
  \langle \ell(p_2,s_2) c(p^\prime,s^\prime)| \;
  \bar{c} \Gamma b
  \; | b(p,s) \ell(p_1,s_1)\rangle_\text{con} \,,
  \label{eq:4q-ME}
\end{align}
where $\Gamma = \gamma_\mu, \gamma_\mu \gamma_5$ are
matrices of the flavor-changing currents, $\ell$ represents a light
spectator quark ($\ell \in \{u,d,s\}$), and $c$ and $b$ represent
charm and bottom quarks, respectively.

\begin{figure*}[t!]
  \subfigure[~One-gluon emission from the $b$ quark]{
    \label{fig:1g-cont-b}
    \includegraphics[width=0.45\textwidth]{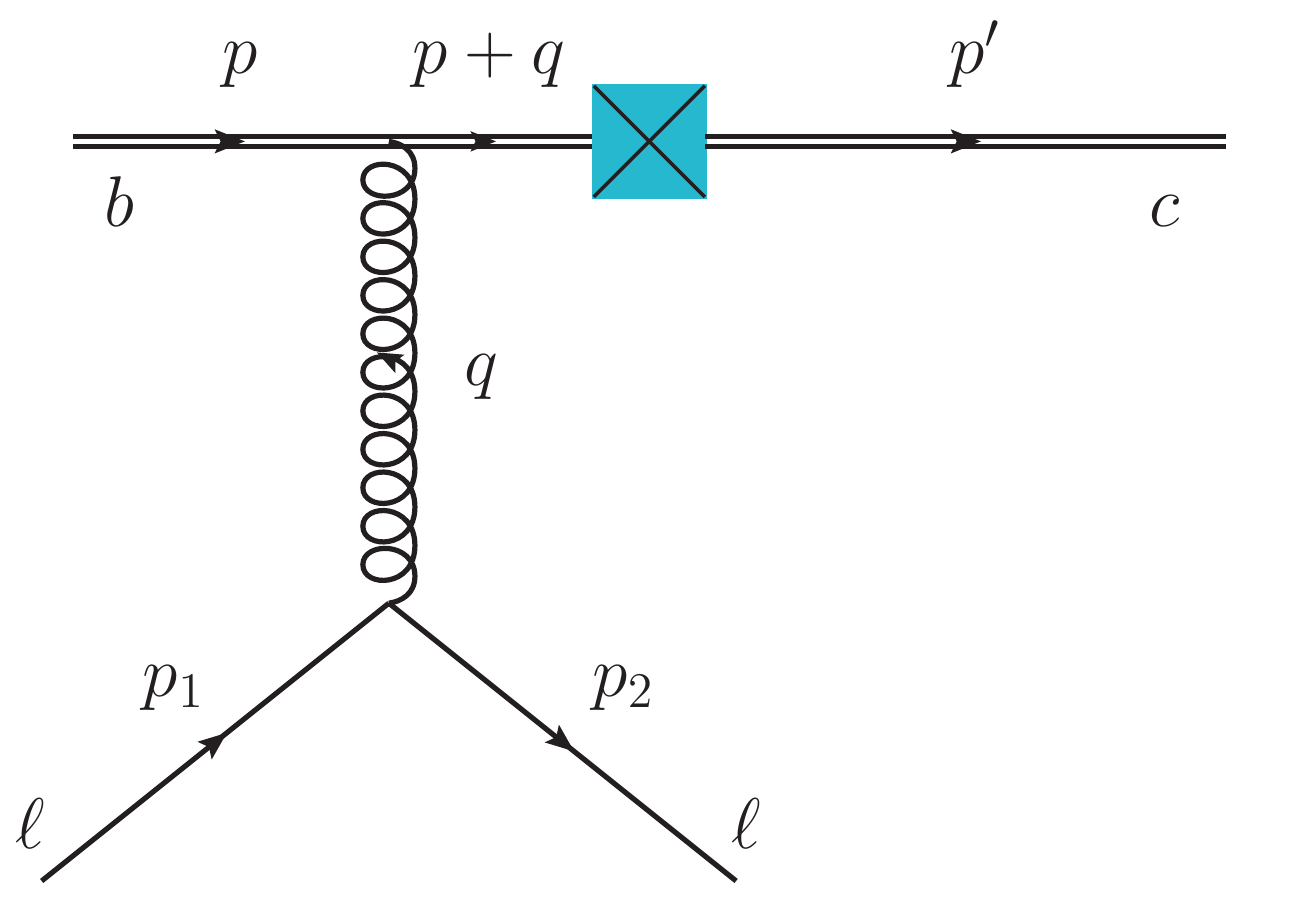}
  }
  \hfill
  \subfigure[~One-gluon emission from the $c$ quark]{
    \label{fig:1g-cont-c}
    \includegraphics[width=0.45\textwidth]{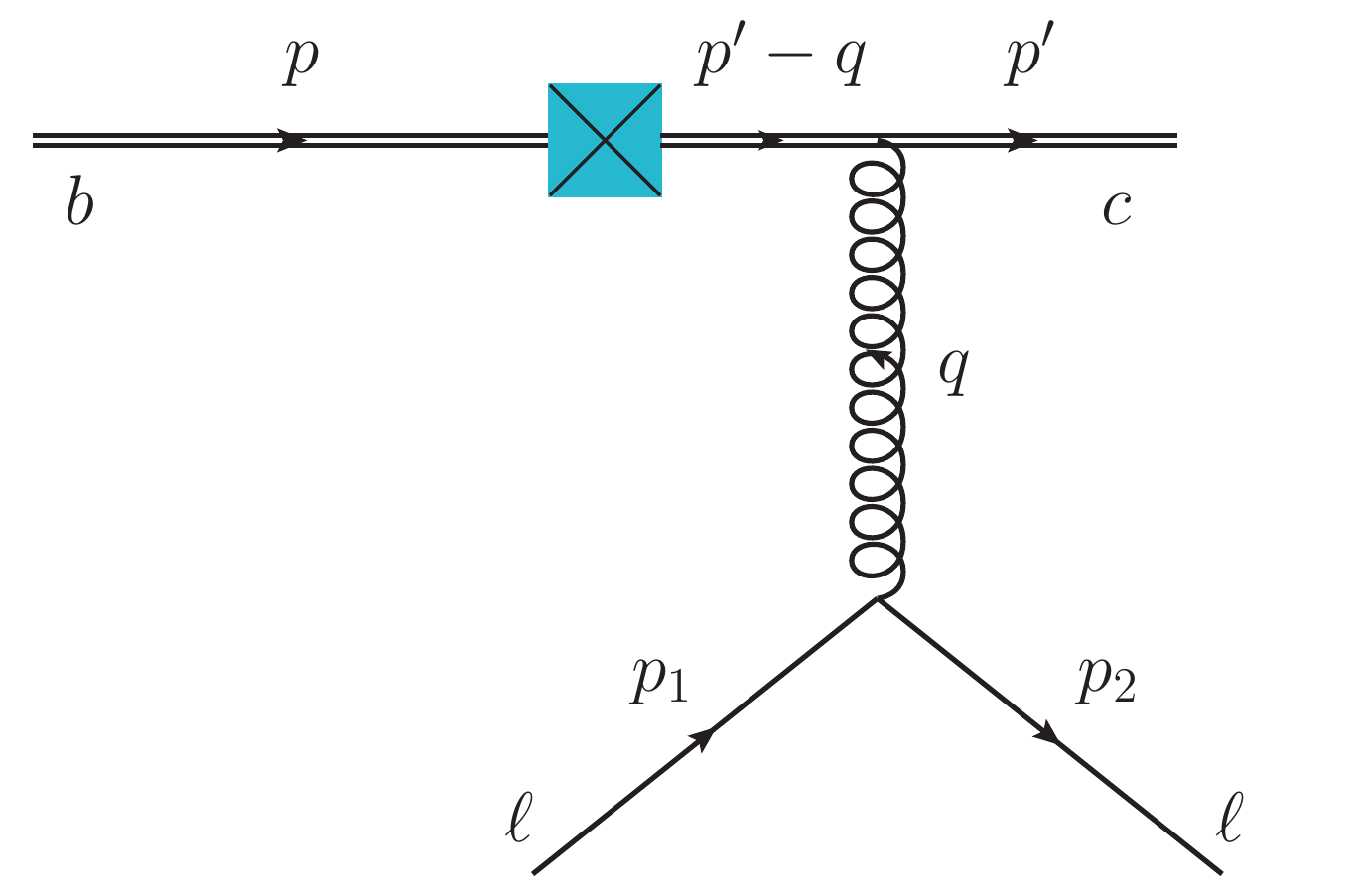}
  }
  \caption{ 
Tree-level continuum diagrams with a gluon exchange.  A colored
box represents an insertion of the flavor-changing operator.
}
\label{fig:cont-diagram}
\end{figure*}

\begin{figure*}[t!]
\label{fig:latt-diagram}
  \subfigure[~One-gluon emission from the action vertex]{
    \label{fig:1g-act}
    \includegraphics[width=0.45\textwidth]{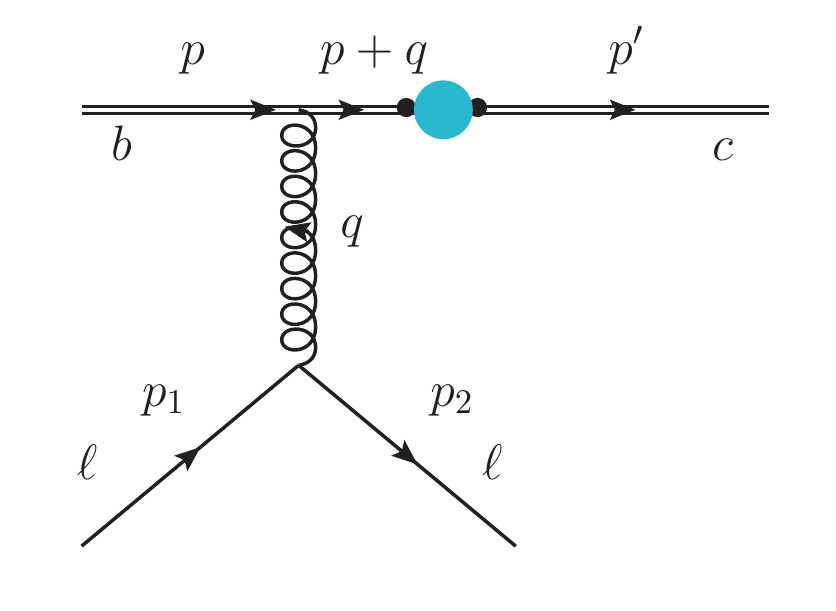}
  }
  \hfill
  \subfigure[~One-gluon emission from the improved quark field ]{
    \label{fig:1g-imp}
    \includegraphics[width=0.45\textwidth]{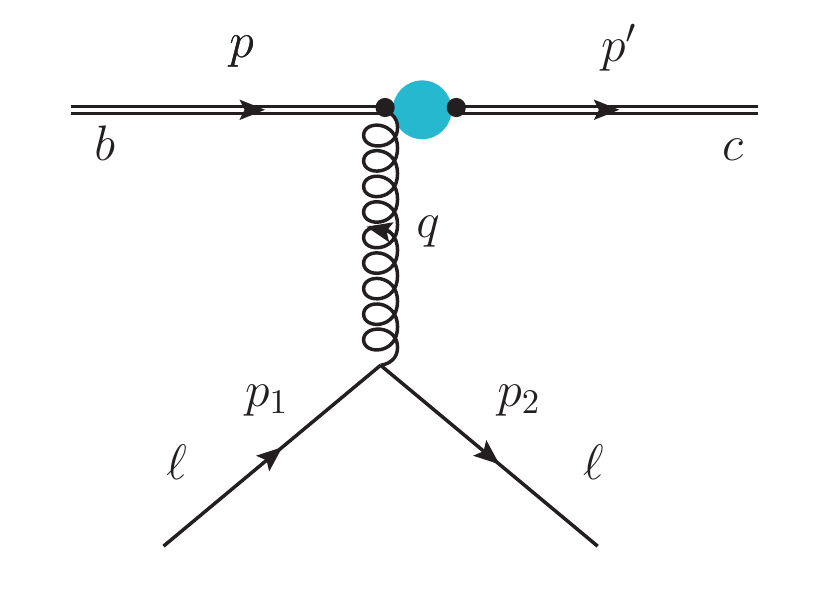}
  }
  \caption{ Tree-level lattice diagrams with one-gluon exchange at the
    $b$-quark line.  A colored circle represents an insertion of the
    flavor-changing current operator.  The black dot without a gluon
    line in \subref{fig:1g-act} and in \subref{fig:1g-imp} represents
    the zero-gluon vertex from the improved quark fields.  The 
    black dot with a gluon line in \subref{fig:1g-imp} represents the
    one-gluon emission vertex from the improved quark field.}
  \label{fig:1g-all}
\end{figure*}

At tree level, the connected diagram contains one-gluon exchange
between the light spectator quark and the heavy quarks.
Here we consider only the diagram with one-gluon exchange at the
$b$-quark line, shown in Fig.~\ref{fig:1g-cont-b}.
The diagram with one-gluon exchange on the $c$-quark line, shown in
Fig.~\ref{fig:1g-cont-c}, is identical if we switch $b \to c$.
The lattice diagrams which correspond to the continuum diagram in
Fig.~\ref{fig:1g-cont-b} are shown in Figs.~\ref{fig:1g-act}
and~\ref{fig:1g-imp}.
One-gluon emission may occur through the one-gluon vertex of the OK
action as in Fig.~\ref{fig:1g-act} or through the vertex of the
improved quark field as in Fig.~\ref{fig:1g-imp}.
The small black dot attached to the current operator (cyan circle) with
(without) a gluon line represents the one-gluon (zero-gluon) vertex of
the improved quark fields.
The charm quark part has a separate matching factor which is
completely factorized from the bottom quark part.

Hence, let us focus on matching the lattice diagrams with one-gluon
exchange on the $b$-quark line in Fig.~\ref{fig:1g-all} to the
continuum diagram in Fig.~\ref{fig:1g-cont-b}.
The matching condition is
\begin{align} 
n_\mu(q) 
\Big[ &R^{(0)}_b(p+q)  S^{\text{lat}}_b(p+q)(-gt^a) \Lambda_\mu(p+q,p) 
\nonumber \\ 
&+(-gt^a)R^{(1)}_{b\mu}(p+q,p)\Big] 
\mathcal{N}_b(p)u^{\text{lat}}_b(p,s) 
\nonumber \\ 
&= 
%\frac{m_b-i\gamma \cdot(p+q)}{m_b^2+(p+q)^2} 
S_b(p+q) (-gt^a)\gamma_\mu  
\sqrt{\frac{m_b}{E_b}}u_b(p,s), 
\label{eq:matching-subdiagram} 
\end{align} 
where $q$ is a four-momentum of the emitted gluon, $\mu$ is a Lorentz
index, and $t^a$ is a generator of the SU(3) color group.
$n_\mu(q)=2\sin(\frac{1}{2}q_\mu)/q_\mu$ is the gluon line
wave-function factor~\cite{Weisz:1982zw}.
$S_b$ and $S^{\text{lat}}_b$ are fermion propagators of $b$ quarks in
the continuum and on the lattice, respectively.
Here $\Lambda_\mu$ is one-gluon emission vertex from the OK action
for $b$ quarks.
$R^{(0)}_b$ and $R^{(1)}_{b,\mu}$ come from the improved quark field
for $b$ quarks.
$R^{(1)}_{b,\mu}$ represents the one-gluon emission vertex from
the improved quark field for $b$ quarks.
Explicit formulas for $\Lambda_\mu$ and $R^{(1)}_{b,\mu}$ are given 
in Appendix \ref{app:LFR-1}.

Both the spatial momentum of the external $b$ quark, $\boldsymbol{p}$,
and the four-momentum of the exchanged gluon, $q$, are $ \mathcal{O}
(\Lambda_{\text{QCD}})$: $\boldsymbol{p}, \boldsymbol{q}, q_4
  \approx \Lambda_{\text{QCD}}$.
They are much smaller than the physical $b$-quark mass, $m_b$, and
the lattice cut-off scale $1/a \cong 1.6\sim 4.5 \GeV$.
Hence, it is possible to expand both sides of
Eq.~\eqref{eq:matching-subdiagram} in power series of
$q/m_b$, $\boldsymbol{p}/m_b$, $qa$, and $\boldsymbol{p}a$.

When we expand in $q$ and $\boldsymbol{p}$ on both sides of
Eq.~\eqref{eq:matching-subdiagram}, a careful treatment is needed with the
expansion of the heavy quark propagator, since it has pole structure.
For example, in the continuum, the heavy quark propagator with
momentum $p+q$ can be expanded as follows,
\begin{align}
\label{eq:prop-expan-conti}
&S(p+q) 
= \frac{m - i\gamma \cdot (p+q)}{m^2 + (p+q)^2}
\nonumber \\
&= \frac{m(1+\gamma_4)-i\gamma_4(\tilde{p}_4 + q_4)
-i\boldsymbol{\gamma}\cdot(\boldsymbol{p}+\boldsymbol{q})}
{2im(\tilde{p}_4 + q_4)+(\tilde{p}_4 + q_4)^2+(\boldsymbol{p}+\boldsymbol{q})^2}
\nonumber \\
&= 
\frac{1}{i(\tilde{p_4} +q_4)} 
\frac{1+\gamma_4}{2} 
+\bigg[ \frac{1-\gamma_4}{4m}
- \frac{\boldsymbol{\gamma } \cdot (\boldsymbol{p}+\boldsymbol{q})}{2m(\tilde{p}_4+q_4)}
\nonumber \\
&
+\frac{(1+\gamma_4)(\boldsymbol{p}+\boldsymbol{q})^2}{4m(\tilde{p}_4+q_4)^2}
\bigg]
+\cdots,
\end{align}
where $\tilde{p}_4$ is
\begin{align}
\label{on-shell-p4}
\tilde{p}_4 = p_4 - im = i
\Big[\frac{\boldsymbol{p}^2}{2m}
-\frac{(\boldsymbol{p}^2)^2}{8m^2} + \cdots
\Big]\,.
\end{align}
Note that $(\tilde{p_4} +q_4, \boldsymbol{p}+\boldsymbol{q})$ is the 
residual momentum of the internal heavy quark with momentum $p+q$.
If we do the power series expansion as in Eq.~\eqref{eq:prop-expan-conti},
then it is natural to identify each term in the matrix element
in terms of HQET.

Similarly, we can apply the power series expansion to the OK-action
heavy quark propagator \cite{ PhysRevD.78.014504}
\begin{align}
S^{\text{lat}}(p+q) &=\Big[ \mu(p+q) -\cos(p_4+q_4) 
\nonumber \\
&+ i\gamma_4 \sin(p_4+q_4) +i \boldsymbol{\gamma \cdot K}(p+q)\Big]^{-1},
\end{align}
where
\begin{align}
K_i(p) &= \sin(p_i)\big[ \zeta - 2c_2 \hat{\boldsymbol{p}}^2 -c_1 \hat{p}^2_i \big],
\\
\mu(p) &= 1+ m_0 
+ \frac{1}{2}r_s \zeta \hat{\boldsymbol{p}}^2 
+ c_4 \sum_i (\hat{p}_i)^4\,.
\label{eq:mu}
\end{align}
Here $\hat{p}_i= 2\sin(p_i/2)$.
Since $\boldsymbol{p},q_\mu \ll 1/a, m_0$, we can expand the 
lattice propagator as in Eq.~\eqref{eq:prop-expan-conti},
\begin{align}
\label{eq:prop-expan-lattice}
S^{\text{lat}}(p+q) 
&=e^{-m_1}\Big[\frac{1}{i(\tilde{p}^{\text{lat}}_4+q_4)}\frac{1+\gamma_4}{2} 
+ \cdots \Big],
\end{align}
where the ellipsis represents higher order terms.
Here, note that 
\begin{align}
\tilde{p}^{\text{lat}}_4 &= p_4 - im_1 
\nonumber \\
&=i\Big[ \frac{1}{2m_2}\boldsymbol{p}^2 
-\frac{1}{6}w_4 \sum_i p_i^4
- \frac{1}{8m_4^2}
\boldsymbol{p}^4 \Big]+ \cdots,
\end{align}
where $m_2$, $m_4$, and $w_4$
\cite{PhysRevD.78.014504} are functions of the OK action 
coefficients.
Their explicit formulas are given in Appendix \ref{app:coeff}.
In the construction of the OK action, the dispersion relation of the 
heavy quark is already matched to the continuum.
This indicates that $m_2 = m_4 = m$ and $w_4 = 0$, so
$\tilde{p}^{\text{lat}}_4 = \tilde{p}_4$ through
$\mathcal{O}(\boldsymbol{p}^4)$.

The expansions of the external quark spinors are 
introduced in Eq.~\eqref{eq:conti-spinor-expansion} and
Eq.~\eqref{eq:lattice-spinor-expansion}.
Finally, we need to expand the lattice vertices
$\Lambda_{\mu}(p+q,p)$, $R^{(0)}(p+q)$, and $R^{(1)}_\mu(p+q,p)$ in
powers of $\boldsymbol{p}a$ and $qa$.
They are analytic in $\boldsymbol{p}a$ and $qa$, and the expansion is
straightforward.
Comparing both sides of the expansion of the matching condition in
Eq.~\eqref{eq:matching-subdiagram}, we obtain a number of constraint
equations for the OK-action parameters $c_i$ and the
current-improvement parameters $d_i$.
These constraints are sufficient to determine all the improvement parameters
$d_i$ through $\lambda^3$ order, and to put constraints on a subset of the
OK-action parameters $c_i$.
The constraints are consistent with the $c_i$ given in \cite{
PhysRevD.78.014504}.

In the discussion that follows, we identify the terms in the expansion
of the matching condition with contributions from (lattice and
continuum) HQET.
This exercise sheds light on the structure of the matching
calculations and leads naturally to useful cross-checks.
Let us begin with the matching calculation at leading order.
First, let us choose $\mu=4$, the time direction.
Then both sides of Eq.~\eqref{eq:matching-subdiagram} are identical,
\begin{align}
\frac{1}{iP_4} (-gt^a) u(0,s),
\end{align}
where $P_4 = \tilde{p}_4 +q_4$.
In HQET this contribution arises from one-gluon emission from the
one-gluon vertex of the leading-order (LO) Lagrangian:
\begin{align}
\mathcal{L}_0 = \bar{h}^+[-D_4 -m ]h^+.
\end{align}

Second, let us choose the spatial direction $\mu = i$ ($i=1,2,3$).
At leading order the right-hand side (R.H.S.) of
Eq.~\eqref{eq:matching-subdiagram} is
\begin{align}
  &\text{R.H.S.} =
  \nonumber \\
&\Big[\frac{-i(2p_i+q_i) +\epsilon_{ijk}\Sigma_j q_k }{2imP_4}
+ \frac{\gamma_i}{2m} \Big]
(-gt^a) u(0,s) \,.
\label{eq:match-leading-2}
\end{align}
Here the first term proportional to $1/P_4$ in
Eq.~\eqref{eq:match-leading-2} represents gluon emission by the
next-to-leading-order (NLO) Lagrangian:
\begin{align}
  \mathcal{L}_1 =
  \bar{h}^+\bigg[ \frac{1}{2m} \boldsymbol{D}^2 + \frac{i}{2m}
    \boldsymbol{\sigma \cdot B}\bigg] h^+\,.
  \label{eq:cont-HQET-NLO-1}
\end{align}
where the definition of the matrix $\Sigma_i$ is in Appendix
\ref{app:notation}.
The second term in Eq.~\eqref{eq:match-leading-2} represents gluon emission by
the NLO correction term in the FWT field rotation for $b$ quarks in the
flavor-changing current, given in Eq.~\eqref{eq:HQET-FWT-1}.

Now, let us consider the 
left-hand-side (L.H.S.) of Eq.~\eqref{eq:matching-subdiagram} with 
spatial direction $\mu = i$, which corresponds to the lattice part
in the matching condition.
\begin{align}
  &\text{L.H.S.} =
  \nonumber \\
&\Big[\frac{-i (2p_i+q_i) }{2i m_2 P_4}
+\frac{\epsilon_{ijk}\Sigma_jq_k}{2im_B P_4}
+ \frac{\gamma_i}{2m_3} \Big]
(-gt^a) u(0,s),
\label{eq:subdiagram-first-spatial}
\end{align}
where $m_2$ and $m_B$ are the kinetic mass and the chromomagnetic
mass at tree level, respectively:
\begin{align}
\frac{1}{2m_2} &
= \frac{\zeta^2}{m_0(2+m_0)}  + \frac{r_s \zeta }{2(1+m_0)}\,,
\\
\frac{1}{2m_B} 
&= \frac{\zeta^2}{m_0(2+m_0)}  + \frac{c_B \zeta }{2(1+m_0)}\,,
\end{align}
and the coefficient $m_3$ includes a correction from the improved 
current
\begin{align}
\frac{1}{2m_3} &=  \frac{\zeta(1+m_0)}{m_0(2+m_0)}-d_1\,.
\end{align}
The first two terms in Eq.~\eqref{eq:subdiagram-first-spatial} come from
the lattice HQET Lagrangian at NLO:
\begin{align}
\mathcal{L}^{\text{lat}}_1
= \bar{h}^+ \bigg[ \frac{1}{2m_2} \boldsymbol{D}^2
+ \frac{i}{2m_B}\boldsymbol{\sigma \cdot B}\bigg] h^+\,,
\label{eq:lat-HQET-next-to-leading}
\end{align}
which is the lattice version of Eq.~\eqref{eq:cont-HQET-NLO-1}.
The matching condition requires that all the masses equal the
physical mass: $m_2= m_B = m_3 = m$.
Here, $m_2 = m_B = m$ is consistent with the original matching of
  the OK action.
The relation $m_3 = m$ reproduces Eq.~\eqref{eq:d1-match},
%two-quark matrix element matching in the previous subsection,
%
\begin{align}
d_1 = \frac{\zeta(1+m_0)}{m_0 (2+m_0)}- \frac{1}{2m}.
\end{align}

For the expansion through $\lambda^3$ order, the 
full expressions are given in Appendix \ref{app:match-sub}.
The continuum part of the expansion 
(the R.H.S. of Eq.~\eqref{eq:matching-subdiagram}) 
is given in 
Eq.~\eqref{eq:append-temoral-continuum-subdiagram} and
Eq.~\eqref{eq:append-spatial-continuum-subdiagram}.
And the lattice part (the L.H.S. of 
Eq.~\eqref{eq:matching-subdiagram}) is given in
Eq.~\eqref{eq:append-temoral-lattice-subdiagram} and
Eq.~\eqref{eq:append-spatial-lattice-subdiagram}.
The mass parameters $m_i$ and symmetry breaking parameters 
$w_i$ and $dw_i$ in Eq.~\eqref{eq:append-temoral-lattice-subdiagram} 
and Eq.~\eqref{eq:append-spatial-lattice-subdiagram} encapsulate the 
lattice artifacts.
They are functions of the OK-action parameters and the 
improvement parameters $d_i$ of the improved quark field.
The explicit formulas for $m_i, w_i$, and $dw_i$ are given in
Appendix \ref{app:coeff}.
The matching conditions are simply
\begin{align}
m_i = m,
%\quad m_i \in \{m_2, m_b, m_E, m_4, m_{B^\prime}\}, 
\quad w_i=0,
 \qquad dw_i=0 .
\label{eq:match}
\end{align}
As we present in Appendix \ref{app:coeff}, the mass parameters $m_i$
can be classified into two groups.
The first group $M_a \equiv \{m_2, m_b, m_E, m_4, m_{B^\prime}\}$ contains 
the masses to be matched by the action matching.
The second group $M_b \equiv \{ m_3, m_{\alpha E}, \cdots, m_6, m_7\}$ 
contains the masses to be matched by the current matching.
We can classify the matching conditions into 
\begin{align}
m_i &= m, \quad m_i \in M_a, \quad
w_i = 0,  
\quad ~\text{   action}
\label{eq:match-action-result}
\\
m_i &= m, \quad m_i \in M_b, \quad
dw_i = 0.
\quad \text{current}
\label{eq:match-current-result}
\end{align}
The matching conditions of Eq.~\eqref{eq:match-action-result} are 
equivalent to a subset of those for the OK action \cite{PhysRevD.78.014504}
: namely, the dispersion relation and background field interaction.
The matching conditions of Eq.~\eqref{eq:match-current-result} determine
a complete set of current improvement parameters $d_i$ at tree
level.
The explicit formulas for $d_i$ are summarized in
Sec.~\ref{sec:results}.

In Sec.~\ref{sec:hqet-description}, we will interpret
the entire matching procedure in the language of HQET. 
Interpreting Eq.~\eqref{eq:matching-subdiagram} in terms of the
continuum and lattice HQET Feynman rules, we will show how the
matching conditions can be factorized systematically.
%
%

%---------------
% SECTION 5.
%\input{hqet}
%---------------
%
\section{Cross-check by heavy quark effective theory 
\label{sec:hqet-description}}
We have cross-checked the final results presented in Section
\ref{sec:results} in several ways.
First, three researchers (Leem, Bailey, Sunkyu Lee) have done the
calculation, and confirmed them.
Second, when we do the matching calculation, it produces about 150
constraints on the eleven improvement parameters.
The constraints also involve the coefficients in the improvement
terms of the original OK action.
The final results reported here are consistent with all the constraints
as well as the OK action coefficients.
Third, we show that the results are consistent with factorization of
the matching condition in accord with the structure of contributions
from HQET.
Here we explain this third consistency check.

If we use HQET as a stepping stone for matching between continuum QCD
($\leftrightarrow$ continuum HQET) and lattice QCD ($\leftrightarrow$ lattice
HQET), the matching condition given in Eq.~\eqref{eq:4q-ME} can be described
by HQET (lattice HQET).
Especially, the subdiagrams in Eq.~\eqref{eq:matching-subdiagram} can
be described by HQET Feynman rules.
For the continuum, the R.H.S. of
Eq.~\eqref{eq:matching-subdiagram} is 
\begin{align}
  & \text{R.H.S.} =
  \nonumber \\
&\bigg[R^{(1)}_{\text{HQ},\mu}(p+q,p)
+\sum_{n=0}^{\infty} R^{(0)}_{\text{HQ}}(p+q)
\Big(\frac{1}{iP_4}\Lambda^{(0)}_{\text{HQ}}(p+q)\Big)^n
\nonumber \\
&\times \frac{1}{iP_4}
\Lambda^{(1)}_{\text{HQ},\mu}(p+q,p)
\bigg]
(-gt^a)u(0,s),
\label{eq:HQ-diagram-continuum}
\end{align}
where $\Lambda^{(0)}_{\text{HQ}}$ and $\Lambda^{(1)}_{\text{HQ},\mu}$
represent the zero-gluon emission and one-gluon emission vertices,
respectively, which come from the HQET Lagrangian 
in Eq.~\eqref{eq:HQET-lambda-third}.
$R^{(0)}_{\text{HQ}}$ and $R^{(1)}_{\text{HQ},\mu}$ represent the zero-gluon
emission and one-gluon emission vertices, respectively, which come from the
FWT transformation in Eq.~\eqref{eq:HQET-FWT-2} between the QCD quark field
$Q$ and the HQET field $h$.
Here $n$ represents the number of perturbative insertions of higher order
terms in the HQET Lagrangian with no gluon emission.
The spinor $u(0,s) = \gamma_4 u(0,s) = u_v(s)$ can be understood as 
the HQET spinor with $v=(1,\boldsymbol{0})$.
The explicit formulas for $R^{(0)}_{\text{HQ}}$,
$R^{(1)}_{\text{HQ},\mu}$, $\Lambda^{(0)}_{\text{HQ}}$, and
$\Lambda^{(1)}_{\text{HQ},\mu}$ are given in
Eq.~\eqref{eq:HQET-zero-gluon-lagrangian}--\eqref{eq:HQET-one-gluon-operator-spatial}
in Appendix \ref{app:HQET-FR}.

Now let us consider the lattice part.  The L.H.S of
Eq.~\eqref{eq:matching-subdiagram} can be arranged as follows,
\begin{align}
  & \text{L.H.S.} =
  \nonumber \\
  & \bigg[R^{\text{lat},(1)}_{\text{HQ},\mu}(p+q,p)
    +\sum_n R^{\text{lat},(0)}_{\text{HQ}}(p+q)
    \Big(\frac{1}{iP_4}\Lambda^{\text{lat},(0)}_{\text{HQ}}(p+q)\Big)^n
    \nonumber \\
    &
    \times\frac{1}{iP_4}
    \Lambda^{\text{lat},(1)}_{\text{HQ},\mu}(p+q,p)
    \bigg]
  (-gt^a)u(0,s),
\label{eq:HQ-diagram-lattice}
\end{align}
where $\Lambda^{\text{lat},(0)}_{\text{HQ}}$ and
$\Lambda^{\text{lat},(1)}_{\text{HQ},\mu}$ are the lattice counterparts
of $\Lambda^{(0)}_{\text{HQ}}$ and $\Lambda^{(1)}_{\text{HQ},\mu}$.
They can be interpreted as the vertices of the HQET Lagrangian which 
is matched to the lattice action.
We showed $1/m$ terms of this Lagrangian in 
Eq.~\eqref{eq:lat-HQET-next-to-leading}.
At order $1/m^2$, the lattice HQET Lagrangian is expressed 
in terms of a single short-distance coefficient $1/m_E^2$,
\begin{align}
\mathcal{L}_2^{\text{lat}} = \bar{h}^+ \bigg[
\frac{\boldsymbol{D} \cdot \boldsymbol{E}- \boldsymbol{E} \cdot \boldsymbol{D}}{8m_E^2} 
+\frac{i \boldsymbol{\sigma} \cdot 
(\boldsymbol{D}\times \boldsymbol{E} - \boldsymbol{E} \times \boldsymbol{D})}{8m_E^2}
\bigg] h^+.
\label{eq:HQ-lattice-2nd}
\end{align}
As given in \cite{ElKhadra:1996mp} and \cite{PhysRevD.78.014504}, 
the condition $m_E =m$ determines the chromoelectric coefficient 
$c_E$ in the action.
At order $1/m^3$, however, tree-level matching of the 
four-quark matrix elements in Eq.~\eqref{eq:4q-ME} cannot give 
constraints on the two-gluon emission terms.
In \cite{PhysRevD.78.014504}, the full matching of the action up to
$1/m^3$ (or $\mathcal{\lambda}^3$) is presented using the two-gluon
emission vertices in Compton scattering.
The explicit formulas for $\Lambda^{\text{lat},(0)}_{\text{HQ}}$,
$\Lambda^{\text{lat},(1)}_{\text{HQ},\mu}$ are given in
Eq.~\eqref{eq:lat-HQET-zero-gluon-lagrangian}--
\eqref{eq:lat-HQET-one-gluon-lagrangian}. 
They are consistent with the results in \cite{PhysRevD.78.014504}.

In Eq.~\eqref{eq:HQ-diagram-lattice}, $R^{\text{lat},(0)}_{\text{HQ}}$
and $R^{\text{lat},(1)}_{\text{HQ},\mu}$ represent the zero-gluon emission and
one-gluon emission vertices, respectively, which are the lattice counterparts
of $R^{(0)}_{\text{HQ}}$ and $R^{(1)}_{\text{HQ},\mu}$, respectively.
As $R^{(0)}_{\text{HQ}}$ and $R^{(1)}_{\text{HQ},\mu}$ come from the FWT
transformation between the QCD and HQET quark fields
(Eq.~\eqref{eq:HQET-FWT-2}), $R^{\text{lat},(0)}_{\text{HQ}}$ and
$R^{\text{lat},(1)}_{\text{HQ},\mu}$ follow from the relation between the
lattice improved quarks and the HQET quarks, for example, $\Psi_b = \mathbf{U}_b^{\text{lat}}
\cdot h_b$.
We obtain this relation, in turn, from the expression for the improved field
given in Eq.~\eqref{eq:improved-current-lambda-third}.  The explicit formulas
for $R^{\text{lat},(0)}_{\text{HQ}}$ and $R^{\text{lat},(1)}_{\text{HQ},\mu}$
are given in
Eq.~\eqref{eq:lat-HQET-zero-gluon-operator}--\eqref{eq:lat-HQET-one-gluon-operator}.

As a result, the matching condition in
Eq.~\eqref{eq:matching-subdiagram} can be factorized as 
matching of individual building blocks as follows,
\begin{align}
\Lambda^{\text{lat},(0)}_{\text{HQ}}
&= \Lambda^{(0)}_{\text{HQ}} ,
&&
\Lambda^{\text{lat},(1)}_{\text{HQ},\mu}
= \Lambda^{(1)}_{\text{HQ},\mu}
\label{eq:matching-factorize-action}
\\
R^{\text{lat},(0)}_{\text{HQ}}
&= R^{(0)}_{\text{HQ}},
&&
R^{\text{lat},(1)}_{\text{HQ},\mu}
= R^{(1)}_{\text{HQ},\mu}.
\label{eq:matching-factorize-current}
\end{align}
Here Eqs.~\eqref{eq:matching-factorize-action} provide the matching
conditions for the action.
Similarly, Eqs.~\eqref{eq:matching-factorize-current} give the
matching conditions for the improved currents.
We obtain $\mathbf{U}^{\text{lat}}$ as follows,
\begin{align}
\mathbf{U}^{\text{lat}} &=
1- \frac{1}{2m_{3}} \boldsymbol{\gamma \cdot D}
\nonumber \\
&+\frac{1}{4m^2_{\alpha_E}}\boldsymbol{\alpha}\cdot \boldsymbol{E}
+\frac{1}{8m^2_{D_\perp^2}}\boldsymbol{D}^2
+\frac{i}{8m^2_{sB}}\boldsymbol{\Sigma}\cdot \boldsymbol{B}
\nonumber \\
&
-\frac{\{\gamma_4D_4,\boldsymbol{\alpha}\cdot \boldsymbol{E}\}}{8m^3_{\alpha_{EE}}}
-\frac{3\{\boldsymbol{\gamma \cdot D},\boldsymbol{D}^2\}}{32m_{\gamma D D_\perp^2}^3}
-\frac{3\{\boldsymbol{\gamma \cdot D},i\boldsymbol{\Sigma}\cdot \boldsymbol{B}\}}{32m_{5}^3}
\nonumber \\
&
-\frac{\{\boldsymbol{\gamma}\cdot \boldsymbol{D},
\boldsymbol{\alpha}\cdot \boldsymbol{E}\}}{16m_{\alpha_{rE}}^3}
+\frac{[\gamma_4 D_4,\boldsymbol{D}^2]}{16m_{6}^3}
+\frac{[\gamma_4 D_4,i\boldsymbol{\Sigma} \cdot \boldsymbol{B}]}{16m_{7}^3}
\nonumber \\
&
+dw_1 \sum_i \gamma_i D_i^3
+\frac{dw_2}{8} [\boldsymbol{\gamma}\cdot \boldsymbol{D},\boldsymbol{D}^2],
\label{eq:HQET-FWTr-2-lattice}
\end{align}
where the coefficients $m_i \in M_b$ and $dw_i$ are identical to those
in the expanded formulas in
Eq.~\eqref{eq:append-temoral-lattice-subdiagram} and
Eq.~\eqref{eq:append-spatial-lattice-subdiagram}.
Explicit formulas for $m_i$ and $dw_i$ are given in Appendix \ref{app:coeff}.

As a result, the matching relation for the flavor-changing currents
is
%Conversely, from the form of the vertices
%$R^{\text{lat},(0)}_{\text{HQ}}$ and
%$R^{\text{lat},(1)}_{\text{HQ},\mu}$, we can derive the following
%
\begin{align}
\label{eq:HQET-lattice-FWT}
\bar{\Psi}_{Ic}\Gamma \Psi(x)_{Ib}
\doteq \bar{h}_c \bar{\mathbf{U}}_c^{\text{lat}}
\Gamma \mathbf{U}_b^{\text{lat}}
h_b \,.
\end{align}
%
%where
%
%
%
%is a lattice FWT transformation corresponding to the continuum FWT
%transformation in Eq.~\eqref{eq:HQET-FWT-1}.
%
The matching conditions can also be written
\begin{align}
\mathbf{U}^{\text{lat}}_b = \mathbf{U}_b,
\end{align}
where $\mathbf{U}_b$ is defined in Eq.~\eqref{eq:HQET-FWT-2}.
This relation is identical to the matching conditions in
Eq.~\eqref{eq:match-current-result}.

%---------------
% SECTION 6.
%\input{result}
%---------------
%
\section{Results}
\label{sec:results}

The final results for the improvement parameters $d_i$ are
\begin{widetext}

\begin{align}
d_1 & = \frac{\zeta(1+m_0)}{m_0(2+m_0)}-\frac{1}{2m},
\\
d_2 & = 
\frac{2\zeta(1+m_0)}{m_0(2+m_0)}d_1
-\frac{r_s\zeta}{2(1+m_0)}
- \frac{\zeta^2(1+m_0)^2}{m_0^2(2+m_0)^2}
+\frac{1}{4m^2},
\\
d_E &= 
-\frac{2(1+m_0)\zeta}{m_0^2(2+m_0)^2}-\frac{(m_0+1)\zeta c_E}{m_0(2+m_0)}
+\frac{1}{2m^2},
\label{eq:de-ljh}
%\\
%\om{d_B} & \om{= }
%\om{\frac{2\zeta(1+m_0)}{m_0(2+m_0)}d_1}
%\om{-\frac{c_B\zeta}{2(1+m_0)}}
%\om{- \frac{\zeta^2(1+m_0)^2}{m_0^2(2+m_0)^2}}
%\om{+\frac{1}{4m^2},}
\\
d_B &= d_2\,,
%\\
%\om{d_{r_E}} & \om{= \frac{1}{16m_3m^2_{\alpha_E}} +\frac{d_1d_E}{4} - \frac{1}{16m^3},}
\\
d_{r_E} &= \frac{d_1d_E}{4} ,
\\
d_{EE} &= \frac{1+m_0}{(m_0^2+2m_0+2)}\Big[-\frac{1}{4m^3}
+\frac{\zeta(1+m_0)(m_0^2+2m_0+2)}{[m_0(2+m_0)]^3}
+\frac{\zeta c_E(1+m_0)}{[m_0(2+m_0)]^2}
+\frac{(2+2m_0+m_0^2)c_{EE}}{m_0(2+m_0)}\Big],
\\
d_3 &= \frac{3c_1+\zeta/2}{\sinh m_1}  -d_1,
\\
d_4 &= 
\frac{\zeta^3(m_0^3+3m_0^2+5m_0+3)}{2m_0^3(2+m_0)^3}
+\frac{r_s\zeta^2(3m_0^2+6m_0+4)}{4m_0^2(2+m_0)^2}
+\frac{2(1+m_0)c_2}{m_0(2+m_0)}
-\frac{(1+m_0)^2\zeta^2}{2m_0^2(2+m_0)^2}d_1
\nonumber \\
&
-\frac{r_s\zeta}{4(1+m_0)}d_1
+\frac{(1+m_0)\zeta d_2}{2m_0(2+m_0)}-\frac{3}{16m^3},
\\
d_5 &= \frac{d_4}{2},
\\
d_6 &=
\frac{2(1+m_0)}{(m_0^2+2m_0+2)}\Big[ 
\frac{\zeta^2 c_E}{4m_0(2+m_0)} 
-\frac{\zeta c_{EE}(m_0^2+2m_0+2)}{2m_0(1+m_0)(2+m_0)}
-\frac{d_E}{4}\Big(d_1-\frac{2\zeta(1+m_0)}{m_0(2+m_0)} \Big)
-\frac{1}{24m} \Big],
\\
d_7 &= d_6\,.
\label{eq:result_end}
\end{align}
\end{widetext}
Here $m_0$ is a bare quark mass defined in Eq.~\eqref{eq:fermi-act-1}.
For numerical work, the procedure for obtaining $m_0$ from a
hopping parameter $\kappa$ is given in Ref.~\cite{ Bailey:2017xjk}.
Note that $m$ is equal to $m_2$, a kinetic quark mass defined in
Eq.~\eqref{eq:short-distance-mass-action-1}.
The coefficients $c_i$ are parameters for the OK action.
%

%---------------
%\input{limit}
%---------------
Assuming $m_0a \ll 1$, we can cross-check the results 
against those from the Symanzik improvement program.
In Table \ref{tab:param-behav}, we show how the coefficients $c_i$ 
of the OK action and $d_i$ of the current behave in the continuum 
limit $m_0a \to 0$.
% (second column)
%and the static quark limit $am_0 \to \infty$ (third column).
%
Here, we tune $\zeta$ so that $m_1 = m_2$ and do not fix the redundant
coupling $r_s$ to make the comparison clear. 
In Appendix \ref{app:Sym}, we show the Symanzik improvement of
the OK action through $\mathcal{O}(a^2)$.
The $\mathcal{O}(a^2)$ study gives restricted information on $c_i$ 
and $d_i$.
It gives terms to the next-to-leading order for 
$c_B, c_E$, $d_1$ and only the leading order for 
$c_1,c_2,c_3,c_{EE},d_2, d_B, d_E$.
At higher order, it does not give any information.
The results from Symanzik improvement are given in 
Eqs.~\eqref{eq:cb-matching}--\eqref{eq:cee-matching} 
(for $c_1,c_2,c_3$, and $c_{EE}$) and 
Eqs.~\eqref{eq:d1-matching}--\eqref{eq:de-matching} 
(for $d_1,d_2,d_B$, and $d_E$).
They are consistent with the expanded formulas of $c_i$ (the second 
column) and $d_i$ (the fourth column) in Table~\ref{tab:param-behav}.
%
%
%

%---------
% TABLE 1.
%---------
\begin{table*}[]
  \renewcommand{\arraystretch}{1.4}
  \caption{ Behavior of the OK action coefficients $c_i$ 
(second column) and the current improvement parameters $d_i$ (fourth column) 
in the continuum limits. 
Here, $\zeta$ is fixed so that $m_1= m_2$.
 }
  \label{tab:param-behav}
\begin{ruledtabular}
\begin{tabular}{l | c | l |c }
%\begin{tabular}{l | c  | c | c}
  Coeff. & $m_0a \to 0$ ($m_1=m_2$)
&  Coeff. & $m_0a \to 0$ ($m_1=m_2$)
%  & $am_0 \to 0$ ($m_1 \ne m_2$)
%  & $am_0 \to \infty$ ($m_1 \ne m_2$) 
  \\ \hline
  $c_B$ & $r_s$  &$d_1$ & $\frac{1}{4}(1-r_s) +\frac{1}{48}\big(1+3r_s^2\big)m_0a +\mathcal{O}\big((m_0a)^2\big)$  
%& $1$
%& $1$   
  \\  \hline 
$c_E$ & $\frac{1}{2}(1+r_s)+\frac{1}{12}\big(-2-3r_s + 3r_s^2\big)m_0 a$
& $d_2=d_B$ & $\frac{1}{16}\big(1-10r_s +r_s^2 \big)+\frac{1}{96}\big(1+23r_s+27r_s^2 -3r_s^3 \big)m_0a$
%& $1 - \frac{1}{2} m_0a+ \mathcal{O}\big((m_0a)^2 \big)$  
%& $\frac{1}{4} + \frac{1}{m_0a} +\mathcal{O}\Big( \frac{1}{(m_0a)^2} \Big)$
\\
&  $+ \mathcal{O}\big((m_0a)^2\big)$   
& & $+\mathcal{O}\big((m_0a)^2\big)$ 
\\ \hline                                                    
$c_1$ & $-\frac{1}{6} +\frac{1}{12}\big(-1 + 5r_s \big)m_0a +\mathcal{O}\big((m_0a)^2\big)$ 
&$d_E$ & $\frac{1}{48}\big(1-6r_s - 3r_s^2 \big)+\frac{1}{48}\big(-1+2r_s + 3r_s^2 \big)m_0a+ \mathcal{O}\big((m_0a)^2\big)$  
%& $-\frac{1}{6} + \frac{1}{3}m_0a+ \mathcal{O}\big((m_0a)^2\big)$ 
%& $\frac{1}{6} m_0a  - \frac{1}{6}\frac{1}{m_0a}+\mathcal{O}\Big(\frac{1}{(m_0a)^2} \Big)$
\\ \hline                                                                                               
$c_2= c_3$ & $\frac{1}{48}\big(-1-6r_s +3 r_s^2 \big)$ 
&
$d_{r_E}$ & $\frac{1}{768}\big(1-7r_s +3r_s^2 +3 r_s^3 \big) $ 
\\
& 
$+\frac{1}{96}\big( -1 -r_s + 3r_s^2 - 3r_s^3 \big)m_0a  +\mathcal{O}\big((m_0a)^2\big)$ 
& & $+\frac{1}{9216}\big( -11 + 30 r_s + 12r_s^2 - 54r_s^3 - 9r_s^4\big)m_0a+\mathcal{O}\big((m_0a)^2\big)$ 
%$+\frac{1}{9216}\big( -11 + 30 r_s + 12r_s^2 - 54r_s^3 - 9r_s^4\big)m_0a+\mathcal{O}\big((m_0a)^2\big)$ 
%& 
%&  
\\ \hline                                                                                                
$c_4$ & $\frac{3}{8}r_s + \frac{3}{16}\big(r_s - r_s^2 \big)m_0a+\mathcal{O}\big((m_0a)^2\big)$ &
$d_{EE}$ & 
$\frac{1}{384}(-1-r_s-3r_s^2-3 r^3)$
\\
& & &
$\textcolor{black}{+\frac{1}{15360}(-9+80r_s +110 r_s^2+120r_s^3 -45 r^4_s)m_0a + \mathcal{O}\big((m_0a)^2\big)}$
\\ \hline                                                                                                                                       $c_5$ & $\frac{1}{4}r_s + \frac{1}{8}\big(r_s -r_s^2 \big) m_0a  +\mathcal{O}\big((m_0a)^2\big)$  
&
$d_{3}$ & $\frac{1}{4}(-1+5r_s)+ \frac{1}{48}\big(-1-3r_s^2\big)m_0a +\mathcal{O}\big((m_0a)^2 \big)$ 
\\ \hline                                                                                            
$c_{EE}$ & $\frac{1}{96}\big(5 + 6r_s -3r_s^2 \big) $ 
&
$d_{4}=2d_5$ & $\frac{1}{384}\big(5-31r_s+15r_s^2+3r_s^3 \big)$  
\\
& $\textcolor{black}{+\frac{1}{192}\big(1-9r_s +3r_s^2 -3r_s^3\big)m_0a +\mathcal{O}\big((m_0a)^2\big)}$ &
& $+\frac{1}{23040}\big(29+570r_s+360r_s^2 -1170r_s^3 -45r_s^4\big)m_0a+\mathcal{O}\big((m_0a)^2\big)$ 
\\ \hline    
& &$d_{6}=d_7$ & 
$\textcolor{black}{\frac{1}{768} (-11-31r_s-9r_s^2+3 r^3)}$
\\
& & &
$\textcolor{black}{+\frac{1}{7680}(-11+255r_s+235 r^2_s -15r_s^3)m_0a
+\mathcal{O}\big((m_0a)^2\big)}$ 
\end{tabular}
\end{ruledtabular}
\end{table*}

Although the $\mathcal{O}(a^2)$ study gives partial
information on $c_i$ and $d_i$, it helps us to investigate a puzzle
involving $d_E$.
The problem is that our result for $d_E$ given in
Eq.~\eqref{eq:de-ljh} is different from that in Ref.~\cite{
ElKhadra:1996mp}.
The result for $d_E$ in Ref.~\cite{ ElKhadra:1996mp} is
\begin{align}
d_E (\text{FNAL}) &= 
\frac{\zeta(1-c_E)(1+m_0a)}{m_0a(2+m_0a)}
\nonumber \\
&-\frac{\zeta(1+m_0a)}{m_2am_0a(2+m_0a)}+\frac{1}{2m_2^2a^2}\,,
\label{eq:d_E-fnal}
\end{align}
which is obtained for the quarkonium system by working up to order
$v^4$ in the power counting of NRQCD.
Our result for $d_E$ is
\begin{align}
d_E(\text{SWME}) 
=& 
-\frac{2(1+m_0a)\zeta}{m_0^2a^2(2+m_0a)^2}
-\frac{\zeta c_E(1+m_0a)}{m_0a(2+m_0a)}
\nonumber \\
&+\frac{1}{2m^2_2a^2}.
\label{eq:de-ljh-2}
\end{align}
Here, for the comparison, we replace $m$ in Eq.~\eqref{eq:de-ljh}
with $m_2$ without loss of generality.
Taking the continuum limit ($m_0 a \to 0$ and
$|\boldsymbol{p}|/m \ll 1$) of these results gives
\begin{align}
d_E(\text{FNAL}) 
&= \frac{1}{16}\big(3- 2r_s -r_s^2 \big) 
\nonumber \\
+& \frac{1}{48} \big(3-2r_s+3r_s^2 \big) m_0a
+\mathcal{O}\big( (m_0a)^2 \big)\,,
\label{eq:FNAL-expand}
\\
d_E(\text{SWME}) 
&= \frac{1}{48}\big(1-6r_s -3r_s^2 \big) 
\nonumber \\
+&  \frac{1}{48}\big(-1+2r_s +3r_s^2 \big) m_0a
+ \mathcal{O}\big( (m_0a)^2 \big) \,.
\label{eq:SWME-expand}
\end{align}
As we can see, even the leading-order terms of $d_E(\text{FNAL})$ and
$d_E(\text{SWME})$ are different from each other.
Our result for the leading term in $d_E(\text{SWME})$ is consistent
with that from Symanzik improvement, given in
Eq.~\eqref{eq:de-matching}.
We have not found any problem in the derivation of $d_E(\text{FNAL})$
in Ref.~\cite{ElKhadra:1996mp}.
Hence, we do not yet understand the source of the difference between
$d_E(\text{FNAL})$ and $d_E(\text{SWME})$.
However, Andreas Kronfeld, one of the authors of
Ref.~\cite{ElKhadra:1996mp} (FNAL) has derived $d_E$ independently,
following our Feynman diagram method, and produced results consistent
with $d_E(\text{SWME})$ \cite{ASK:2021}.
The Hamiltonian method that produced $d_E(\text{FNAL})$ is under
investigation \cite{ASK:2021}.

Next, let us consider the static limit. 
In the Fermilab method \cite{ElKhadra:1996mp,PhysRevD.78.014504}, 
lattice discretization error is bounded in the static limit. 
If we set the improvement parameters to zero: $d_j = 0$ or the 
action coefficients to zero : $c_j = 0$, the discretization error 
comes from mismatches between lattice mass-like terms 
$m_i(d_j=0,c_j=0)$ and the physical mass $m$, or from pure lattice 
artifacts $w_i$ and $dw_i$.
For example, if one does not introduce the second order improvement 
parameter $d_2$ in the improved current, the discretization error 
propagates from the discrepancy between $1/(8m^2)$ and
$1/(8m_{D_\perp}^2)|_{d_2=0}$ 
(with $d_2$ = 0 in Eq.~\eqref{eq:short-distance-mass-current-d2}).
Likewise, if one does not introduce the chromoelectric term in the action
($c_E = 0$), the discretization error will propagate from the discrepancy 
between $1/(4m^2)$ and $1/(4m_E^2) |_{c_{E}=0}$.
As we can see in Eq.~\eqref{eq:short-distance-mass-action-E}
and Eq.~\eqref{eq:short-distance-mass-current-d2},
$1/4m_E^2 |_{c_{E}=0}$ and $1/(8m_{D_\perp}^2)|_{d_2=0}$ 
behave smoothly as $am_0 \to \infty$.
The other terms of the action matching in 
Eq.~\eqref{eq:short-distance-mass-action-1} -
Eq.~\eqref{eq:short-distance-mass-action-2} 
and those in the current matching in
Eq.~\eqref{eq:short-distance-mass-current-1} -
Eq.~\eqref{eq:short-distance-mass-current-2} 
have the same property.
The smooth behavior makes it possible to control the discretization
errors even for heavy quarks with $m_Q a > 1$.

%---------------
% SECTION 7.
%\input{conclu}
%---------------
\section{Conclusion }
\label{sec:conclusion}

The goal of this paper is to improve the current operators through
$\lambda^3$ order in HQET power counting, the same level
as the OK action.
These improved currents can be used to calculate the semi-leptonic
form factors for the $\BtoDst$, $\BtoD$, $\BtoPi$, and $\BstoK$ decays
and the decay constants $f_B$ and $f_D$.
Our final results for the improvement coefficients $d_i$ are presented
in Section \ref{sec:results}.

We adopt the concept of the improved quark field in Ref.~\cite{
  ElKhadra:1996mp} and extend it to $\mathcal{O}(\lambda^3)$ at tree
level.
We find that one needs to add seven more terms of higher dimension and
corresponding improvement parameters at order $\lambda^3$ 
to Eq.~(A.17) of Ref.~\cite{ ElKhadra:1996mp}.
With one exception (the $d_3$ term), the higher dimension lattice
operators are lattice versions of operators in the continuum FWT
transformation. 
%of the heavy quark field.
%
The $d_3$ operator is required to compensate for
rotation-symmetry-breaking contributions from the normalized spinors
of the OK action.
Thus, we need eleven improvement terms in total.

Our matching conditions in Eq.~\eqref{eq:two-match-subdiagram-1}
and Eq.~\eqref{eq:matching-subdiagram} determine the improvement 
parameters uniquely.
Our final results given in Section \ref{sec:results} have been checked
in several ways.
First, three individuals (Jaehoon Leem, Jon Bailey, Sunkyu Lee) have 
performed the calculation and cross-checked the results against 
one another.
Second, the matching condition provides about 150 self-consistent
constraint equations.
The constraint equations from the temporal and spatial components of
the one-gluon emission vertex are consistent with each other.
The constraint equations from the zero-gluon emission vertex are also
consistent with those from two-quark matrix elements.
As a by-product, the matching condition reproduces the constraint
equations for the zero-gluon and one-gluon emission vertices of the OK
action.
In addition, the matching condition can be expressed in terms of
contributions from continuum HQET 
(Eq.~\eqref{eq:HQ-diagram-continuum})
and lattice HQET (Eq.~\eqref{eq:HQ-diagram-lattice}). 
For the quark-level matrix elements we match, the vertices of the
continuum currents and action are in one-to-one correspondence with
the vertices of the lattice currents and action 
(Eq.~\eqref{eq:matching-factorize-action} and 
Eq.~\eqref{eq:matching-factorize-current}).
This one-to-one mapping provides another cross-check on the final
results in Section \ref{sec:results}.
At the same time, we note that Eq.~\eqref{eq:HQET-lattice-FWT} is
established for the quark-level matrix elements we match by
constructing the rotation matrix from the ansatz for the improved
field.

There remains a puzzle involving $d_E$.
Our result (SWME) is given in Eq.~\eqref{eq:de-ljh-2}.
At present, there is another result (FNAL) for $d_E$ available in 
Ref.~\cite{ElKhadra:1996mp} which is presented in
Eq.~\eqref{eq:d_E-fnal}.  
They are different from each other even at
leading order in the continuum limit.
To check the validity of our result, we have performed Symanzik
improvement assuming $m_0a \ll 1$ and $|\boldsymbol{p}|/m \ll 1$.
We find the result is consistent with our result for $d_E$.
However, we have not found any problem with the derivation of $d_E$ in
Ref.~\cite{ElKhadra:1996mp}.
Therefore, this issue needs further investigation.

\vspace*{5mm}

\begin{acknowledgments}
  The research of W.~Lee is supported by the Mid-Career Research
  Program (Grant No.~NRF-2019R1A2C2085685) of the NRF grant funded by
  the Korean government (MOE).
  This work was supported by Seoul National University Research Grant in
  2019.
  W.~Lee would like to acknowledge the support from the KISTI
  supercomputing center through the strategic support program for the
  supercomputing application research (No.~KSC-2016-C3-0072,
  KSC-2017-G2-0009, KSC-2017-G2-0014, KSC-2018-G2-0004,
  KSC-2018-CHA-0010, KSC-2018-CHA-0043, KSC-2020-CHA-0001).
  Computations were carried out in part on the DAVID clusters at
  Seoul National University.
\end{acknowledgments}

\appendix

%------------------
% APPENDIX A.
%\input{appendix}
%------------------
%
\section{Notation}
\label{app:notation}

We use the same signature for the $\gamma$-matrices as in
Ref.~\cite{ ElKhadra:1996mp}.
The representation for Euclidean gamma matrices is
\begin{align}
\boldsymbol{\gamma} = 
\begin{pmatrix}
~0 & \boldsymbol{\sigma}~ \\
~\boldsymbol{\sigma} & 0~
\end{pmatrix},
\qquad
\gamma_4 = 
\begin{pmatrix}
~1 & 0 ~\\
~0 & -1~
\end{pmatrix},
\end{align}
where $\boldsymbol{\sigma}$ are Pauli matrices.
The $\gamma$-matrices satisfy the Clifford algebra:
\begin{align}
  \{\gamma_\mu, \gamma_\nu\} &= 2 \delta_{\mu\nu}
\end{align}
The remaining definitions are
\begin{align}
\boldsymbol{\alpha} 
= 
\begin{pmatrix}
~0 & \boldsymbol{\sigma}   \\
-\boldsymbol{\sigma} & 0~
\end{pmatrix},
\qquad
\boldsymbol{\Sigma} =
\begin{pmatrix}
~\boldsymbol{\sigma} & 0~  \\
~0 & \boldsymbol{\sigma} ~
\end{pmatrix},
\end{align}
where $\alpha_i = \gamma_4 \gamma_i$ and 
$\Sigma_k  = -\frac{i}{4} \epsilon_{ijk}[\gamma_i, \gamma_j]$.

\begin{widetext}
  \section{Matching sub-diagrams}
  \label{app:match-sub}
The expansion of the right-hand side (continuum) of
Eq.~\eqref{eq:matching-subdiagram} through third order in $\lambda$ is
as follows,
\begin{align}
\label{eq:append-temoral-continuum-subdiagram}
\text{R.H.S }&(\mu=4)=
\Bigg[
\frac{1}{iP_4}
-\frac{\boldsymbol{\gamma \cdot (p+q)}}{2m P_4}
+\frac{(\boldsymbol{p}+\boldsymbol{q})^2}{2mP_4^2}
-\frac{i\boldsymbol{\gamma \cdot q}}{4m^2}
+\epsilon_{ijk}\Sigma_i \frac{q_j p_k}{4m^2 P_4}
+\frac{i\big(\boldsymbol{ p}^2
+ 2\boldsymbol{q} \cdot (\boldsymbol{p}+\boldsymbol{q})\big)}{8m^2 P_4}
-\frac{i\boldsymbol{\gamma} \cdot (\boldsymbol{p}+\boldsymbol{q})
(\boldsymbol{p}+\boldsymbol{q})^2}{4m^2 P_4^2}
\nonumber \\
&
+\frac{i((\boldsymbol{p}+\boldsymbol{q})^2)^2}{4m^2 P_4^3}
-\frac{\boldsymbol{q} \cdot(\boldsymbol{p}+\boldsymbol{q})}{8m^3}
+\frac{\boldsymbol{\gamma} \cdot \boldsymbol{q}}{8m^3}q_4
+\epsilon_{ijk}\Sigma_i \frac{iq_j p_k}{8m^3}
+\frac{\boldsymbol{\gamma} \cdot (\boldsymbol{p}+\boldsymbol{q})\boldsymbol{p}^2}{16m^3 P_4}
+\frac{\boldsymbol{\gamma} \cdot (\boldsymbol{p}+2\boldsymbol{q})
(\boldsymbol{p}+\boldsymbol{q})^2}{8m^3 P_4}
\nonumber \\
&
-\frac{(\boldsymbol{p}+\boldsymbol{q})^2 \big(3(\boldsymbol{p}+\boldsymbol{q})^2
+\boldsymbol{q}^2\big)}{16m^3 P_4^2}
+\epsilon_{ijk}\Sigma_i \frac{iq_j p_k (\boldsymbol{p}+\boldsymbol{q})^2}{8m^3 P_4^2}
+\frac{\boldsymbol{\gamma} \cdot (\boldsymbol{p}+\boldsymbol{q})
\big((\boldsymbol{p}+\boldsymbol{q})^2\big)^2}{8m^3P_4^3}
-\frac{((\boldsymbol{p}+\boldsymbol{q})^2)^3}{8m^3 P_4^4}\Bigg]
(-gt^a) u(0,s),
\\
\text{R.H.S }&(\mu=i)=
\Bigg[
\frac{1}{2m}\gamma_i
-\frac{(2p_i + q_i) +\epsilon_{ijk}i\Sigma_j q_k}{2m P_4}
% NLO
+\frac{iq_4}{4m^2}\gamma_i
-\frac{i(p_i+q_i)}{4m^2}
+ \epsilon_{ijk}\Sigma_j \frac{q_k+p_k}{4m^2}
+\frac{i(\boldsymbol{p}+\boldsymbol{q})\cdot \boldsymbol{q}}{4m^2 P_4}\gamma_i
+\frac{i \boldsymbol{\gamma} \cdot (\boldsymbol{p}+\boldsymbol{q}) p_i}{4m^2 P_4}
\nonumber \\
&
+\frac{i\boldsymbol{\gamma} \cdot \boldsymbol{p}(p_i+q_i)}{4m^2 P_4}
-\epsilon_{ijk}\gamma_5 \frac{i q_j p_k}{4m^2 P_4}
+ \epsilon_{ijk}
\frac{\Sigma_j q_k(\boldsymbol{p}+\boldsymbol{q})^2}{4m^2 P_4^2}
-i(2p_i+q_i)
\frac{(\boldsymbol{p}+\boldsymbol{q})^2}{4m^2 P_4^2}
% NNLO
+\frac{(p_i+q_i)q_4}{8m^3}
-\frac{q_4^2}{8m^3}\gamma_i
-\frac{\boldsymbol{p}^2}{8m^3}\gamma_i
\nonumber \\
&-\frac{3(\boldsymbol{p}+\boldsymbol{q})^2 +\boldsymbol{q}^2}{16m^3}\gamma_i
-\frac{p_i\boldsymbol{\gamma}\cdot(\boldsymbol{p}+\boldsymbol{q})}{8m^3}
-\frac{(p_i+q_i)\boldsymbol{\gamma}\cdot \boldsymbol{p}}{8m^3}
+\epsilon_{ijk}\Sigma_j \frac{i(p_k+q_k)q_4}{8m^3}
+\epsilon_{ijk} \gamma_5 \frac{q_j p_k}{8m^3}
+\frac{(4p_i+q_i)\boldsymbol{p}^2}{16m^3 P_4}
\nonumber \\
&
+\frac{(3p_i+2q_i)(\boldsymbol{p}+\boldsymbol{q})^2}{8m^3 P_4}
+\epsilon_{ijk}\Sigma_j \frac{i(q_k -2p_k)\boldsymbol{p}^2}
{16m^3 P_4}
+\epsilon_{ijk}\Sigma_j \frac{i(p_k+2q_k)(\boldsymbol{p}+\boldsymbol{q})^2}
{8m^3 P_4}
-\frac{p_i \boldsymbol{\gamma \cdot}(\boldsymbol{p}+\boldsymbol{q})
(\boldsymbol{p}+\boldsymbol{q})^2}{8m^3 P_4^2}
\nonumber \\
&
-\frac{(p_i+q_i) \boldsymbol{\gamma \cdot}\boldsymbol{p}
(\boldsymbol{p}+\boldsymbol{q})^2}{8m^3 P_4^2}
-\frac{\boldsymbol{q}\cdot(\boldsymbol{p}+\boldsymbol{q})(\boldsymbol{p}+\boldsymbol{q})^2}
{8m^3 P_4^2}\gamma_i
+\epsilon_{ijk}\gamma_5 \frac{q_j p_k (\boldsymbol{p}+\boldsymbol{q})^2}
{8m^3 P_4^2}
+\epsilon_{ijk}\Sigma_j \frac{iq_k \big((\boldsymbol{p}+\boldsymbol{q})^2\big)^2}
{8m^3 P_4^3}
\nonumber \\
&
+\frac{(2p_i+q_i)\big((\boldsymbol{p}+\boldsymbol{q})^2\big)^2}
{8m^3 P_4^3}\Bigg] (-gt^a)u(0,s).
\label{eq:append-spatial-continuum-subdiagram}
\end{align}
And the left-hand side (lattice) is
\begin{align}
\label{eq:append-temoral-lattice-subdiagram}
\text{L.H.S }&(\mu=4)=
\Bigg[
\frac{1}{iP_4}
-\frac{\boldsymbol{\gamma \cdot (p+q)}}{2m_3P_4}
+\frac{(\boldsymbol{p}+\boldsymbol{q})^2}{2m_2 P_4^2}
-\frac{i \boldsymbol{\gamma} \cdot \boldsymbol{q}}{4m^2_{\alpha_E}}
+\frac{i(\boldsymbol{p}+\boldsymbol{q})^2}{8m_{D^2_\perp}^2P_4}
+\frac{i \boldsymbol{q}^2+2\epsilon_{ijk}\Sigma_i  q_j p_k
}{8m_E^2 P_4}
-\frac{i\boldsymbol{\gamma}\cdot (\boldsymbol{p}+\boldsymbol{q})
(\boldsymbol{p}+\boldsymbol{q})^2}{4m_2m_3 P_4^2}
\nonumber \\
&
+\frac{i\big((\boldsymbol{p}+\boldsymbol{q})^2\big)^2}{4m^2_2 P_4^3}
-\frac{\boldsymbol{q}^2-2 i \epsilon_{ijk} \Sigma_i q_j p_k}{16m^3_{\alpha_{r E}}}
-\frac{\boldsymbol{q}\cdot (2\boldsymbol{p}+\boldsymbol{q})}{16m_6^3}
+\frac{\boldsymbol{\gamma }\cdot \boldsymbol{q}
}{8m^3_{\alpha_{EE}}} q_4
+\frac{\boldsymbol{ \gamma} \cdot (\boldsymbol{p}+\boldsymbol{q})\boldsymbol{p}^2
-\boldsymbol{\gamma} \cdot ({\boldsymbol{p}-\boldsymbol{q})}(\boldsymbol{p}+\boldsymbol{q})^2}{16 m_3 m_E^2 P_4}
\nonumber \\
&
+\frac{3\boldsymbol{\gamma} \cdot (\boldsymbol{p}+\boldsymbol{q})(\boldsymbol{p}+\boldsymbol{q})^2}
{16m^3_{\gamma D D_\perp^2}P_4}
-\frac{(\boldsymbol{p}+\boldsymbol{q})^2\big(\boldsymbol{q}^2 -2i\epsilon_{ijk}\Sigma_i q_j p_k \big)}{16m_2m_E^2 P_4^2}
-\frac{\big((\boldsymbol{p}+\boldsymbol{q})^2\big)^2}{16m_2 m^2_{D^2_\perp}P_4^2}
-\frac{\big((\boldsymbol{p}+\boldsymbol{q})^2\big)^2}{8m_4^3 P_4^2}
-\frac{dw_1}{6P_4}\sum_i \gamma_i (p_i +q_i)^3
\nonumber \\
&
-\frac{w_4}{6P_4^2}\sum_i (p_i+q_i)^4
+\frac{ \boldsymbol{\gamma} \cdot (\boldsymbol{p}+\boldsymbol{q})}{8m_2^2m_3 P_4^3}
\big((\boldsymbol{p}+\boldsymbol{q})^2\big)^2
-\frac{\big((\boldsymbol{p}+\boldsymbol{q})^2\big)^3}{8m^3_2 P_4^4}
\Bigg]
(-gt^a)u(0,s),
\\
\text{L.H.S }&(\mu=i)=
\Bigg[
\frac{1}{2m_3}\gamma_i
-\frac{(2p_i + q_i)}{2m_2 P_4}
-\frac{i\epsilon_{ijk}\Sigma_j q_k}{2m_B P_4}
+\frac{iq_4}{4m_{\alpha_E}^2} \gamma_i
-i\frac{2p_i+q_i}{8m_{D^2_\perp}^2}
+ \epsilon_{ijk}\Sigma_j
\frac{q_k}{8m_{sB}^2}
-\frac{i \big(q_i + i\epsilon_{ijk}\Sigma_j (2p_k +q_k) \big)}{8m_E^2}
\nonumber \\
&
+\frac{i (\boldsymbol{p}+\boldsymbol{q})\cdot \boldsymbol{q}\gamma_i}{4m_3 m_B P_4}
+\frac{i(2p_i+q_i)(\boldsymbol{p}+\boldsymbol{q})\cdot \boldsymbol{\gamma}}{4m_3m_2 P_4}
-\frac{i(p_i+q_i)\boldsymbol{q}\cdot \boldsymbol{\gamma}}{4m_3m_BP_4}
-i \epsilon_{ijk}\frac{q_jp_k \gamma_5}{4m_3m_B P_4}
+\epsilon_{ijk}\Sigma_j \frac{q_k (\boldsymbol{p}+\boldsymbol{q})^2}{4m_Bm_2 P_4^2}
\nonumber \\
&
-\frac{i(2p_i+q_i)(\boldsymbol{p}+\boldsymbol{q})^2}{4m_2^2 P_4^2}
-\frac{q_4^2}{8m^3_{\alpha_{EE}}}\gamma_i
+\frac{q_4(2p_i+q_i)}{16m_{6}^3}
+\frac{q_iq_4}{16m^3_{\alpha_{rE}}}
-\frac{3\big((\boldsymbol{p}+\boldsymbol{q})^2+\boldsymbol{p}^2\big)}{32m_{\gamma D D^2_\perp}^3}\gamma_i
% +d_{z_3}\boldsymbol{\gamma \cdot q}q_i  %%corr
-\frac{3\boldsymbol{q}^2}{32m_{5}^3}\gamma_i
\nonumber \\
&-\frac{(\boldsymbol{p}+\boldsymbol{q})\cdot (2\boldsymbol{p}+\boldsymbol{q})}{16m_3m_E^2}\gamma_i
+\frac{3q_i \boldsymbol{\gamma}\cdot \boldsymbol{q}}{32m_{5}^3}
-\frac{3(2p_i+q_i)\boldsymbol{\gamma}\cdot (2\boldsymbol{p}+\boldsymbol{q})}
{32m_{\gamma D D_\perp^2}^3}
+\frac{(2p_i+q_i)\boldsymbol{\gamma}\cdot \boldsymbol{p}
+p_i\boldsymbol{\gamma}\cdot\boldsymbol{q}}{16m_3m_E^2}
+ i \epsilon_{ijk} \Sigma_j \frac{q_4q_k}{16m_7^3}
\nonumber \\
&
%-d_{z_3} \boldsymbol{q}^2 \gamma_i
+ i \epsilon_{ijk} \Sigma_j \frac{q_4 (2p_k+q_k)}{16m_{\alpha_{rE}}^3}
+\epsilon_{ijk}\gamma_5
\frac{3q_j p_k}{16m_{5}^3}
-\epsilon_{ijk}\gamma_5 \frac{q_jp_k}{16m_E^2m_3}
+\frac{dw_2}{8} \big(
-\boldsymbol{q}\cdot (2\boldsymbol{p}+\boldsymbol{q}) \gamma_i
+\boldsymbol{\gamma}\cdot \boldsymbol{q}(2p_i+q_i)
\big)
\nonumber \\
&
+\frac{1}{6}dw_1 \gamma_i (3p_i^2+3p_iq_i+q_i^2)
+\frac{(2p_i + q_i)((\boldsymbol{p}+\boldsymbol{q})^2 +\boldsymbol{p}^2)}{8m_4^3 P_4}
+\frac{\boldsymbol{q}\cdot(2\boldsymbol{p}+\boldsymbol{q}) q_i}{16m_2m_E^2 P_4}
+\frac{(2p_i + q_i)(\boldsymbol{p}+\boldsymbol{q})^2}{16m_{D_\perp^2}^2m_2 P_4}
\nonumber \\
&
+i \epsilon_{ijk}\Sigma_j \frac{q_k (\boldsymbol{p}^2+(\boldsymbol{p}+\boldsymbol{q})^2)}{8m_{B^\prime}^3 P_4}
+i \epsilon_{ijk}\Sigma_j \frac{q_k (\boldsymbol{p}+\boldsymbol{q})^2}{16m_{D^2_\perp}^2 m_B P_4}
+i \epsilon_{ijk}\Sigma_j
\frac{(2p_k+q_k) \big(\boldsymbol{q}\cdot (2\boldsymbol{p}+\boldsymbol{q})\big)}{16m_2m^2_E P_4}
-\frac{w_B}{8P_4}(p_i \boldsymbol{q}^2 -q_i \boldsymbol{p}\cdot \boldsymbol{q})
\nonumber \\
&
-\frac{w_B}{16P_4}i\epsilon_{ijk}\Sigma_j q_k \boldsymbol{q}^2
+\frac{w_4}{6P_4}(2p_i+q_i)\big((p_i+q_i)^2+p_i^2\big)
+\frac{w_B}{8P_4}i\epsilon_{ijk}q_j p_k \boldsymbol{\Sigma}\cdot(2\boldsymbol{p}+\boldsymbol{q})
+\frac{w^\prime_B}{12P_4}i \epsilon_{ijk}\Sigma_j q_k (q_i^2 + q_k^2)
\nonumber \\
&
+\frac{(w_4+w^\prime_4)}{12P_4}i\epsilon_{ijk} \Sigma_j q_k \Big((3p_i^2 + 3p_i q_i + q_i^2)
+ (3p_k^2 + 3p_k q_k+ q_k^2) \Big)
-\frac{(2p_i+q_i)(\boldsymbol{p}+\boldsymbol{q})\cdot \boldsymbol{\gamma}
(\boldsymbol{p}+\boldsymbol{q})^2}{8m_3m_2^2 P_4^2}
\nonumber \\
&
+\frac{(p_i+q_i)\boldsymbol{q}\cdot \boldsymbol{\gamma}
(\boldsymbol{p}+\boldsymbol{q})^2}{8m_3m_2 m_B P_4^2}
-\frac{\boldsymbol{q}\cdot (\boldsymbol{p}+\boldsymbol{q})
(\boldsymbol{p}+\boldsymbol{q})^2\gamma_i}{8m_3m_2m_B P_4^2}
+\epsilon_{ijk}\gamma_5 \frac{q_j p_k (\boldsymbol{p}+\boldsymbol{q})^2}{8m_3m_2m_B
P_4^2}
+i \epsilon_{ijk}\Sigma_j
\frac{q_k ((\boldsymbol{p}+\boldsymbol{q})^2)^2}{8m_2^2m_B P_4^3}
\nonumber \\
&
+\frac{(2p_i+q_i)\big((\boldsymbol{p}+\boldsymbol{q})^2\big)^2}{8m_2^3 P_4^3}
\Bigg](-gt^a)u(0,s).
\label{eq:append-spatial-lattice-subdiagram}
\end{align}

\end{widetext}

\begin{widetext}
  %-----------------
  % APPENDIX D.
  %\input{appendix2}
  %-----------------
  %
  \section{Lattice Feynman rules}
  \label{app:LFR-1}
The one-gluon vertices of the OK action from Ref.~\cite{
  PhysRevD.78.014504} are as follows (set $a=1$),
\begin{align}
\label{eq:one-gluon-OK-action}
\Lambda_4(p+q,p)
 = \gamma_4 \cos \big( p_4+\frac{1}{2}q_4 \big)
&- i\sin \big(p_4+\frac{1}{2} q_4 \big)
+ \frac{i}{2}c_E \zeta \sum_i \alpha_i \sin q_i
\cos \frac{1}{2}q_4
\nonumber \\
&+ c_{EE} \sum_i\gamma_i \cdot \sin q_i \big[\sin(p+q)_4 - \sin p_4 \big]
\cos \frac{1}{2}q_4,
\end{align}
\begin{align}
\Lambda_i(p+q,p)
&= \zeta \gamma_i \cos \big(p_i +\frac{1}{2}q_i \big)
-ir_s \zeta \sin \big(p_i +\frac{1}{2}q_i)
-\frac{i}{2}c_E \zeta  \alpha_i \sin q_4\cos \frac{1}{2}q_i
-\frac{1}{2}c_B \zeta  \epsilon_{irm}\Sigma_m \sin q_r
\cos \frac{1}{2}q_i
\nonumber \\
&
-c_2 
\bigg[\gamma_i \cos(p_i +\frac{1}{2}q_i)\sum_j 4\big[ \sin^2 \frac{1}{2}(p_j+q_j)
+\sin^2 \frac{1}{2}p_j \big]
+2\sin(p_i +\frac{1}{2}q_i)\sum_j \gamma_j \big[\sin(p_j +q_j) +\sin p_j \big]
\bigg]
\nonumber \\
&-\frac{1}{2}c_1 \gamma_i \bigg[4\cos(p_i +\frac{1}{2}q_i)
\big[ \sin^2 \frac{1}{2}(p_i+q_i) +\sin^2 \frac{1}{2}p_i \big]
+2\sin(p_i +\frac{1}{2}q_i) \big[\sin(p_i +q_i) +\sin p_i \big]
\bigg]
\nonumber \\
&+c_3\cos \frac{1}{2}q_i 
\bigg[\sum_j \gamma_j \sin q_j \big[ \sin(p_i +q_i) - \sin p_i \big]
-\gamma_i \sum_j \sin q_j \big[\sin(p_j+q_j)-\sin p_j\big]
\nonumber \\
&
\qquad\qquad\qquad
-\gamma_4 \gamma_5 \sum_{r,m}\epsilon_{irm} \sin q_r \big[\sin(p_m+q_m) +\sin p_m \big]\bigg]
\nonumber \\
&-c_{EE}\gamma_i \sin q_4\big[ \sin (p_4+q_4) -\sin p_4 \big]\cos \frac{1}{2}q_i
-8ic_4\sin(p_i +\frac{1}{2}q_i)\big[\sin^2\frac{1}{2}(p_i +q_i)+\sin^2\frac{1}{2}p_i \big]
\nonumber \\
&
-4c_5 \epsilon_{irm}\Sigma_m \sin q_r
\cos \frac{1}{2}q_i
\Big[
\sum_j\big[ \sin^2 \frac{1}{2}\big(p_j +q_j)+\sin^2 \frac{1}{2}p_j \big]
-\big[ \sin^2 \frac{1}{2}\big(p_m +q_m\big)+\sin^2 \frac{1}{2}p_m \big]
\Big]
\,.
\end{align}

The zero-gluon vertex of the improved quark field is as follows,
\begin{align}
\label{eq:zero-gluon-impr-field}
R^{(0)}(p+q)
&=e^{m_1/2} \Bigg[
1+id_1 \sum_j \gamma_j \sin(p_j +q_j) - 2d_2\sum_j \sin^2 \frac{1}{2}(p_j+q_j)
\nonumber \\
&-\frac{2}{3}i d_3 \sum_j \gamma_j \sin(p_j+q_j)\sin^2 \frac{1}{2}(p_j+q_j)
-4i d_4 \sum_{j,k}\gamma_j \sin(p_j+q_j) \sin^2 \frac{1}{2}(p_k+q_k)
\Bigg]\,.
\end{align}

The one-gluon vertices of the improved quark field are as follows,
\begin{align}
\label{eq:one-gluon-impr-field}
R^{(1)}_{4} (p+q,p)
=e^{m_1/2}&\cos \frac{1}{2}q_4 \gamma_4 \Bigg[
\frac{i}{2}d_E  \sum_j \gamma_j \sin q_j
-d_{EE}\gamma_4 \sum_j \gamma_j \sin q_j
\big[\sin(p_4 +q_4) - \sin p_4\big]
\nonumber \\
&+d_{r_E}\sum_j \sin q_j \big[\sin(p_j+q_j)-\sin p_j \big]
-id_{r_E}\sum_{j,l,m} \epsilon_{jlm} \Sigma_j \sin q_l \big[\sin(p_m+q_m) +\sin p_m\big]
\Bigg]
\nonumber \\
-4e^{m_1/2}& d_6 \cos(p_4+\frac{1}{2}q_4) \gamma_4
        \sum_j \big[\sin^2 \frac{1}{2}(p_j+q_j) -\sin^2 \frac{1}{2}p_j \big]
\,,
\end{align}

\begin{align}
R^{(1)}_{i}(p+q,p)
=e^{m_1/2}& \Bigg[-d_1  \gamma_i \cos (p_i +\frac{1}{2}q_i)
-id_2 \sin (p_i +\frac{1}{2}q_i)
-\frac{1}{2}d_B \sum_{r,m} \epsilon_{irm} \Sigma_m \sin q_r \cos \frac{1}{2}q_i
\nonumber \\
&
-\frac{i}{2}d_E \gamma_4 \gamma_i \cos \frac{1}{2}q_i \sin q_4
+d_{r_E} \sum_{r,m}i\epsilon_{irm}\Sigma_m \gamma_4 \sin q_4 \big[\sin (p_r+q_r)+\sin p_r \big]
\cos \frac{1}{2}q_i
\nonumber \\
&
-d_{r_E}\gamma_4 \sin q_4 \big[\sin (p_i+q_i)-\sin p_i \big] \cos \frac{1}{2}q_i
+d_{EE}\gamma_i \sin q_4
\big[ \sin(p_4 +q_4) -\sin p_4 \big]\cos \frac{1}{2}q_i
\nonumber \\
&+\frac{1}{2}d_4
\bigg[\gamma_i \cos(p_i +\frac{1}{2}q_i)\sum_j 4\big[ \sin^2 \frac{1}{2}(p_j+q_j)
+\sin^2 \frac{1}{2}p_j \big]
+2\sin(p_i +\frac{1}{2}q_i)\sum_j \gamma_j \big[\sin(p_j +q_j) +\sin p_j \big]
\bigg]
\nonumber \\
&+\frac{1}{12}d_3 \gamma_i \bigg[4\cos(p_i +\frac{1}{2}q_i)
\big[ \sin^2 \frac{1}{2}(p_i+q_i) +\sin^2 \frac{1}{2}p_i \big]
+2\sin(p_i +\frac{1}{2}q_i) \big[\sin(p_i +q_i) +\sin p_i \big]
\bigg]
\nonumber \\
&+d_5 \cos \frac{1}{2}q_i
\Big[-\sum_j \gamma_j \sin q_j \big[ \sin(p_i +q_i) - \sin p_i \big]
+\gamma_i \sum_j \sin q_j \big[\sin(p_j+q_j)-\sin p_j\big]\Big]
\nonumber \\
&+d_5 \cos \frac{1}{2}q_i \gamma_4 \gamma_5 \sum_{r,m}\epsilon_{irm}
\sin q_r \big[\sin(p_m+q_m) +\sin p_m \big]
\nonumber \\
&+2d_6 \gamma_4 \big[\sin(p_4+q_4)-\sin p_4 \big] \sin(p_i +\frac{1}{2}q_i)
\nonumber \\
&-i\sum_{r,m}\epsilon_{irm}d_7 \big[\sin(p_4+q_4)-\sin p_4 \big] 
\cos \frac{1}{2}q_i \sin q_r \Sigma_m \gamma_4\Bigg]\,.
\end{align}
The factor for the external incoming fermion with momentum $p$ and
spin $s$ is given by $\mathcal{N}(p)u^{\text{lat}}(p,s)$ with the
normalization factor $\mathcal{N}(p)$ and the spinor
$u^{\text{lat}}(p,s)$ as follows \cite{ ElKhadra:1996mp,
  PhysRevD.78.014504},
\begin{align}
\mathcal{N}(p) 
&= \bigg( 
\frac{\mu(p)-\cosh E }{\mu(p)\sinh E}
\bigg)^{1/2},
\\
u^{\text{lat}}(p,s)
&= \frac{\mu(p)-\cosh E + \sinh E - i\boldsymbol{\gamma \cdot K}}
{\sqrt{2 (\mu(p)- \cosh E)(\mu(p) -\cosh E +\sinh E)}}
u(0,s),
\end{align}
where $\mu(p)$ is given in Eq.~\eqref{eq:mu} and $u(0,s)$ is a constant
spinor which satisfies $\gamma_4 u(0,s) = u(0,s)$.
Here, $\mathcal{N}(p)$ corresponds to $\sqrt{\frac{m}{E}}$ and
$u^{\text{lat}}(p,s)$ corresponds to the continuum spinor as follows,
\begin{align}
u(p,s) = \frac{m + E -i \boldsymbol{\gamma \cdot p}}{\sqrt{2m(m+E)}} u(0,s).
\end{align}

\section{HQET Feynman rules}
\label{app:HQET-FR}
The zero-gluon vertex of the HQET Lagrangian is as follows,
\begin{align}
\label{eq:HQET-zero-gluon-lagrangian}
\Lambda^{(0)}_{\text{HQ}}(p) = -\frac{1}{2m}\boldsymbol{p}^2
+\frac{1}{8m^3}\big(\boldsymbol{p}^2\big)^2.
\end{align}
The one-gluon vertices of the HQET Lagrangian are as follows,
\begin{align}
\label{eq:HQET-one-gluon-lagrangian}
\Lambda^{(1)}_{\text{HQ},4}(p+q,p)
&=\Big[
 1- \frac{\boldsymbol{q}^2 - 2i \epsilon_{ijk} q_i p_j \Sigma_k}{8m^2} \Big],
\\
\Lambda^{(1)}_{\text{HQ},i}(p+q,p)
&= \Big[
-\frac{i}{2m}(2p_i+q_i)
+\frac{1}{2m}\epsilon_{ijk} \Sigma_j q_k
+\frac{q_4}{8m^2}\big(q_i+ i\epsilon_{ijk}\Sigma_j (2p_k+q_k)\big)
+\frac{i(2p_i+q_i)}{8m^3}
\big((\boldsymbol{p}+\boldsymbol{q})^2+\boldsymbol{p}^2\big)
\nonumber \\
& 
\qquad
-\frac{1}{8m^3}\epsilon_{ijk}\Sigma_j q_k
\big((\boldsymbol{p}+\boldsymbol{q})^2+\boldsymbol{p}^2\big)
\Big],
\end{align}

The zero-gluon vertex from Eq.~\eqref{eq:HQET-FWT-2} is as follows,
\begin{align}
\label{eq:HQET-zero-gluon-operator}
R^{(0)}_{\text{HQ}}(p) =
1&-\frac{i}{2m}\boldsymbol{\gamma}\cdot \boldsymbol{p}
-\frac{1}{8m^2}\boldsymbol{p}^2
+\frac{3i\boldsymbol{\gamma}\cdot \boldsymbol{p}}{16m^3}\boldsymbol{p}^2.
\end{align}
The one-gluon vertices from Eq.~\eqref{eq:HQET-FWT-2} are as follows,
\begin{align}
\label{eq:HQET-one-gluon-operator}
R^{(1)}_{\text{HQ},4}(p+q,p) &=
-\frac{i}{4m^2}\boldsymbol{\gamma}\cdot \boldsymbol{q}
+\frac{q_4}{8m^3}\boldsymbol{\gamma}\cdot \boldsymbol{q}
-\frac{1}{16m^3}\big(\boldsymbol{q}^2 -2i\epsilon_{ijk}\Sigma_i q_jp_k\big)
-\frac{1}{16m^3}\big(\boldsymbol{q}^2 + 2\boldsymbol{p}\cdot \boldsymbol{q} \big),
\end{align}
\begin{align}
R^{(1)}_{\text{HQ},i}(p+q,p) &=
\frac{1}{2m}\gamma_i
+\frac{i}{4m^2}q_4 \gamma_i-\frac{i}{8m^2}(2p_i+q_i)
+\frac{1}{8m^2}\epsilon_{ijk}\Sigma_j q_k
-\frac{q_4^2}{8m^3}\gamma_i
\nonumber \\
&
-\frac{3}{32m^3}\big(
\boldsymbol{\gamma}\cdot (2\boldsymbol{p}+\boldsymbol{q})(2p_i+q_i)
+(\boldsymbol{p}^2+(\boldsymbol{p}+\boldsymbol{q})^2)\gamma_i\big)
-\frac{3}{32m^3}i\epsilon_{ijk}q_k \big(\Sigma_j \boldsymbol{\gamma}\cdot\boldsymbol{p}
+\boldsymbol{\gamma}\cdot(\boldsymbol{p}+\boldsymbol{q})\Sigma_j \big)
\nonumber \\
&+\frac{q_4}{16m^3}\big(i\epsilon_{ijk}\Sigma_j(2p_k+q_k) +q_i\big)
+\frac{q_4}{16m^3}(2p_i+q_i)
+\frac{q_4}{16m^3}i\epsilon_{ijk}\Sigma_j q_k.
\label{eq:HQET-one-gluon-operator-spatial}
\end{align}

The zero-gluon vertex of the lattice HQET Lagrangian is as follows,
\begin{align}
\label{eq:lat-HQET-zero-gluon-lagrangian}
\Lambda^{\text{lat},(0)}_{\text{HQ}}(p) = -\frac{1}{2m_2}\boldsymbol{p}^2
+\frac{1}{8m_4^3}\big(\boldsymbol{p}^2\big)^2
+\frac{1}{6}w_4 \sum_i p_i^4.
\end{align}
The one-gluon vertices of the lattice HQET Lagrangian are as follows,
\begin{align}
\Lambda^{\text{lat},(1)}_{4,\text{HQ}}(p+q,p)
&=\Big[
1- \frac{\boldsymbol{q}^2 - 2i \epsilon_{ijk} q_i p_j \Sigma_k}{8m_E^2} \Big],
\\
\Lambda^{\text{lat},(1)}_{i,\text{HQ}}(p+q,p)
&= \Big[
-\frac{i}{2m_2}(2p_i+q_i)
+\frac{1}{2m_B}\epsilon_{ijk} \Sigma_j q_k
+\frac{q_4}{8m_E^2}\big(q_i+ i\epsilon_{ijk}\Sigma_j (2p_k+q_k)\big)
+\frac{i(2p_i+q_i)}{8m_4^3}
\big((\boldsymbol{p}+\boldsymbol{q})^2+\boldsymbol{p}^2\big)
\nonumber \\
&
-\frac{1}{8m_{B^{\prime}}^3}\epsilon_{ijk}\Sigma_j q_k
\big((\boldsymbol{p}+\boldsymbol{q})^2+\boldsymbol{p}^2\big)
+\frac{i}{6}w_4 (2p_i +q_i)\big((p_i+q_i)^2+p_i^2\big)
-\frac{i}{8}w_{B_1}\big(p_i \boldsymbol{q}^2 - q_i \boldsymbol{p}\cdot
\boldsymbol{q}\big)
\nonumber \\
&
-\frac{1}{16}w_{B_2}\epsilon_{ijk}\Sigma_j q_k \boldsymbol{q}^2
-\frac{1}{8}w_{B_3}\epsilon_{ijk}q_j p_k \boldsymbol{\Sigma}\cdot(2\boldsymbol{p}+\boldsymbol{q})
-\frac{1}{12}w^\prime_B \epsilon_{ijk}\Sigma_j q_k (q_i^2 + q_k^2)
\nonumber \\
&
-\frac{1}{12}(w_4+w^\prime_4)\epsilon_{ijk} \Sigma_j q_k \Big((3p_i^2 + 3p_i q_i + q_i^2)
+ (3p_k^2 + 3p_k q_k+ q_k^2) \Big)
\Big].
\label{eq:lat-HQET-one-gluon-lagrangian}
\end{align}

The zero-gluon vertex from Eq.~\eqref{eq:HQET-FWTr-2-lattice} is as
follows,
\begin{align}
\label{eq:lat-HQET-zero-gluon-operator}
R^{\text{lat},(0)}_{\text{HQ}}(p) =
1&-\frac{i}{2m_3}\boldsymbol{\gamma}\cdot \boldsymbol{p}
-\frac{1}{8m^2_{D^2_\perp}}\boldsymbol{p}^2
+\frac{3i\boldsymbol{\gamma}\cdot \boldsymbol{p}}{16m^3_{\gamma D D_\perp^2}}\boldsymbol{p}^2
-dw_1 \sum_j i\gamma_j p^3_j,
\end{align}
The one-gluon vertices from Eq.~\eqref{eq:HQET-FWTr-2-lattice} are as follows,
\begin{align}
R^{\text{lat},(1)}_{\text{HQ},4}(p+q,p) &=
-\frac{i}{4m_{\alpha E}^2}\boldsymbol{\gamma}\cdot \boldsymbol{q}
+\frac{q_4}{8m_{\alpha_{EE}}^3}\boldsymbol{\gamma}\cdot \boldsymbol{q}
-\frac{1}{16m_{\alpha_{rE}}^3}\big(\boldsymbol{q}^2 -2i\epsilon_{ijk}\Sigma_i q_jp_k\big)
-\frac{1}{16m_{6}^3}\big(\boldsymbol{q}^2 + 2\boldsymbol{p}\cdot \boldsymbol{q} \big),
\\
R^{(1)}_{\text{HQ},i}(p+q,p) &=
\frac{1}{2m_3}\gamma_i
+\frac{i q_4}{4m_{\alpha E}^2}\gamma_i
-\frac{i}{8m_{D^2_\perp}^2}(2p_i+q_i)
+\frac{\epsilon_{ijk}\Sigma_j q_k}{8m_{sB}^2}
-\frac{q_4^2}{8m_{\alpha_{EE}}^3}\gamma_i
\nonumber \\
&-\frac{3}{32m_{\gamma D D_\perp^2}^3}\big(
\boldsymbol{\gamma}\cdot (2\boldsymbol{p}+\boldsymbol{q})(2p_i+q_i)
+(\boldsymbol{p}^2+(\boldsymbol{p}+\boldsymbol{q})^2)\gamma_i\big)
-\frac{3i\epsilon_{ijk}q_k }{32m^3_5}
\big(\Sigma_j \boldsymbol{\gamma}\cdot\boldsymbol{p}
+\boldsymbol{\gamma}\cdot(\boldsymbol{p}+\boldsymbol{q})\Sigma_j \big)
\nonumber \\
&
+\frac{q_4}{16m_{\alpha_{rE}}^3}\big(i\epsilon_{ijk}\Sigma_j(2p_k+q_k) +q_i\big)
+\frac{q_4}{16m_6^3}(2p_i+q_i)
+\frac{q_4}{16m_7}i\epsilon_{ijk}\Sigma_j q_k
 +dw_1 \gamma_i(3p_i^2 +3p_i q_i+q_i^2)
\nonumber\\
&
+\frac{1}{8}dw_2\big(\boldsymbol{q}\cdot (2\boldsymbol{p}+\boldsymbol{q}) \gamma_i
+\boldsymbol{\gamma}\cdot \boldsymbol{q}(2p_i+q_i)\big).
\label{eq:lat-HQET-one-gluon-operator}
\end{align}

%--------------
% APPENDIX F
%--------------
\section{Short-distance coefficients}
\label{app:coeff}
The lattice short-distance coefficients which determine the action
coefficients are as follows (set $a=1$),
\begin{align}
\label{eq:short-distance-mass-action-1}
\frac{1}{2m_2} &
= \frac{\zeta^2}{m_0(2+m_0)}  + \frac{r_s \zeta }{2(1+m_0)},
\qquad
\frac{1}{2m_B}
= \frac{\zeta^2}{m_0(2+m_0)}  + \frac{c_B \zeta }{2(1+m_0)},
\\
\label{eq:short-distance-mass-action-E}
\frac{1}{4m_E^2}
&= \frac{\zeta^2}{m_0^2(2+m_0)^2} + \frac{\zeta^2 c_E}{m_0(2+m_0)},
\\
\frac{1}{m_4^3}
& = \frac{8\zeta^4}{m_0^3(2+m_0)^3}
+\frac{4\zeta^4 +8r_s\zeta^3 (1+m_0)}{m_0^2(2+m_0)^2}
+\frac{r_s^2\zeta^2}{(1+m_0)^2}
+ \frac{32\zeta c_2}{m_0(2+m_0)},
\\
\frac{1}{m_{B^\prime}^3}
&= \frac{1}{m_4^3} - \frac{r_s(r_s-c_B)\zeta^2}{(1+m_0)^2},
\\
w_B
&= \frac{4(r_s-c_B)\zeta^3(1+m_0)}{m_0^2(2+m_0)^2}+\frac{16\zeta(c_2-c_3)}{m_0(2+m_0)},
\qquad
w^\prime_B
 = \frac{c_B\zeta -4c_5}{1+m_0},
\\
w_4 &=\frac{2\zeta(\zeta+6c_1)}{m_0(2+m_0)} +\frac{r_s\zeta -24c_4}{4(1+m_0)},
\qquad
w^\prime_4
 = -\frac{r_s \zeta -24c_4+32c_5}{4(1+m_0)}.
\label{eq:short-distance-mass-action-2}
\end{align}

The lattice short-distance coefficients which determine the improvement
parameters are as follows (set $a=1$),
\begin{align}
\label{eq:short-distance-mass-current-1}
\frac{1}{2m_3} &=  \frac{\zeta(1+m_0)}{m_0(2+m_0)}-d_1,
\\
\frac{1}{4m^2_{\alpha E}}
&=
\frac{(1+m_0)\zeta}{m_0^2(2+m_0)^2}+\frac{(m_0+1)\zeta c_E}{2m_0(2+m_0)}
+\frac{d_E}{2},
\\
\frac{1}{8m^2_{D_\perp^2}}
&=
-\frac{\zeta(1+m_0)}{m_0(2+m_0)}d_1
+\frac{r_s\zeta}{4(1+m_0)}
+ \frac{\zeta^2(1+m_0)^2}{2m_0^2(2+m_0)^2}
+\frac{d_2}{2},
\label{eq:short-distance-mass-current-d2}
\\
\frac{1}{8m^2_{sB}}
&=
-\frac{\zeta(1+m_0)}{m_0(2+m_0)}d_1
+\frac{c_B\zeta}{4(1+m_0)}
+ \frac{\zeta^2(1+m_0)^2}{2m_0^2(2+m_0)^2}
+\frac{d_B}{2},
\\
\frac{1}{16m_{\alpha_{rE}}^3}
&
= \frac{1}{16m_3m^2_{\alpha_E}} +\frac{d_1d_E}{4} -d_{r_E},
\\
\frac{1}{16m_{\alpha_{EE}}^3} &= \frac{(1+m_0)(m_0^2+2m_0+2)\zeta}{4m_0^3(2+m_0)^3}
+\frac{(1+m_0)\zeta c_E}{4m_0^2(2+m_0)^2}
\nonumber \\
&+\frac{(m_0^2+2m_0+2)c_{EE}}{4m_0(2+m_0)}-\frac{(m_0^2+2m_0+2)d_{EE}}{4(1+m_0)},
\\
\frac{3}{16m_{\gamma D D_\perp^2}^3} &=
\frac{\zeta^3(m_0^3+3m_0^2+5m_0+3)}{2m_0^3(2+m_0)^3}
+\frac{r_s\zeta^2(3m_0^2+6m_0+4)}{4m_0^2(2+m_0)^2}
+\frac{2(1+m_0)c_2}{m_0(2+m_0)}
\nonumber \\
&
-\frac{(1+m_0)^2\zeta^2}{2m_0^2(2+m_0)^2}d_1
-\frac{r_s\zeta}{4(1+m_0)}d_1
+\frac{(1+m_0)\zeta d_2}{2m_0(2+m_0)}-d_4,
\\
\frac{3}{16m_{5}^3} &=
\frac{\zeta^3(m_0^3+3m_0^2+5m_0+3)}{2m_0^3(2+m_0)^3}
+\frac{c_B\zeta^2(3m_0^2+6m_0+4)}{4m_0^2(2+m_0)^2}
+\frac{2(1+m_0)c_3}{m_0(2+m_0)}
\nonumber \\
&
-\frac{(1+m_0)^2\zeta^2}{2m_0^2(2+m_0)^2}d_1
-\frac{c_B\zeta}{4(1+m_0)}d_1
+\frac{(1+m_0)\zeta d_B}{2m_0(2+m_0)}-2d_5,
\\
\frac{1}{16m_{6}^3} &= \frac{1}{16m_3m^2_{\alpha E}} -\frac{\zeta^2 c_E}{4m_0(2+m_0)}
+\frac{\zeta c_{EE}(m_0^2+2m_0+2)}{2m_0(1+m_0)(2+m_0)}
\nonumber \\
&+\frac{d_E}{4}\Big(d_1-\frac{2\zeta(1+m_0)}{m_0(2+m_0)} \Big)
+\frac{1}{24m_2}
+\frac{(m_0^2+2m_0+2)}{2(1+m_0)}d_6,
\\
\frac{1}{16m_{7}^3}
&= \frac{1}{16m_3m^2_{\alpha E}} -\frac{\zeta^2 c_E}{4m_0(2+m_0)}
+\frac{\zeta c_{EE}(m_0^2+2m_0+2)}{2m_0(1+m_0)(2+m_0)}
\nonumber \\
&+\frac{d_E}{4}\Big(d_1-\frac{2\zeta(1+m_0)}{m_0(2+m_0)} \Big)
+\frac{1}{24m_B}
+\frac{(m_0^2+2m_0+2)}{2(1+m_0)}d_7,
\\
dw_1 &= d_3 +d_1 - \frac{3c_1+\zeta/2}{\sinh m_1}, 
\\
dw_2 &= \frac{\zeta(r_s - c_B) }{1+m_0}d_1
+ \frac{\zeta^2(r_s-c_B)+2\zeta (d_2-d_B)(1+m_0)}{m_0(2+m_0)}.
\label{eq:short-distance-mass-current-2}
\end{align}

%-----------------
% APPENDIX F
%\input{appendix3}
%-----------------
\section{Symanzik improvement program ($m_0a \to 0$ limit)}
\label{app:Sym}
In this section we consider the improvement of the action and current
in the limit $m_0a \to 0$ through $\mathcal{O}(a^2)$.
In doing so we reproduce the leading-order behavior of the action and
current improvement parameters in Table \ref{tab:param-behav}.
In the $m_0a \to 0$ limit, one can expand the OK action in $a$,
%
%The OK action through the second order in $a$ is given by,
%
\begin{align}
\label{eq:Wilson-dim-4}
S_{\text{OK},a^2} = 
\sum_x a^4 ~ \bar{\psi}(x)\bigg[
&m_0 +  \gamma_4D_{\text{lat},4} 
+\zeta \boldsymbol{ \gamma }\cdot \boldsymbol{D}_{\text{lat}}
-\frac{1}{2}a \Delta_4 
-\frac{1}{2}r_s \zeta a   \Delta^{(3)}  
\nonumber \\
&-\frac{1}{2}c_B \zeta a  i 
\boldsymbol{\Sigma} \cdot \boldsymbol{B}_{\text{lat}}\psi(x)
-\frac{1}{2}c_E \zeta a 
\boldsymbol{\alpha} \cdot \boldsymbol{E}_{\text{lat}}\psi(x)
\nonumber \\
& +  c_1 a^2
\sum_i \gamma_i D_{\text{lat},i}\Delta_{\text{lat},i} 
+ c_2 a^2 \{  \boldsymbol{\gamma} \cdot \boldsymbol{D}_{\text{lat}}, 
\Delta^{(3)}\} \nonumber \\
& + c_3 a^2 \{ \boldsymbol{\gamma} \cdot \boldsymbol{D}_{\text{lat}},
i \boldsymbol{\Sigma} \cdot \boldsymbol{B}_{\text{lat}}\}
+ c_{EE} a^2 \{\gamma_4 D_{\text{lat},4},
\boldsymbol{\alpha} \cdot \boldsymbol{E}_{\text{lat}}\}
\bigg]
\psi(x).
\end{align}
The corresponding local effective Lagrangian through 
$\mathcal{O}(a^2)$ is given by
\begin{align}
S_{\text{Sym}}  
= \int d^4 x ~ \bar{\psi}(x)\bigg[
&m_0 +  \Big(\gamma_4D_{4} 
+\frac{1}{6} \gamma_4 a^2 D_4^3
\Big)
+\zeta \Big(
\boldsymbol{ \gamma }\cdot \boldsymbol{D}
+\frac{1}{6}  \sum_{i} \gamma_i a^2 D_i^3
\Big)
\nonumber \\
&
-\frac{1}{2}a
D_4^2 
-\frac{1}{2} r_s \zeta
a \boldsymbol{D}^2
-\frac{1}{2} c_B \zeta   i 
 a\boldsymbol{\Sigma} \cdot \boldsymbol{B} 
-\frac{1}{2} c_E \zeta 
a \boldsymbol{\alpha} \cdot \boldsymbol{E} 
\nonumber \\
&+  c_1  
\sum_i \gamma_i a^2 D_i^3
+ c_2 a^2 \{  \boldsymbol{\gamma} \cdot \boldsymbol{D}, 
\boldsymbol{D}^2 \}
+ c_3 a^2 \{ \boldsymbol{\gamma} \cdot \boldsymbol{D} ,
i \boldsymbol{\Sigma} \cdot \boldsymbol{B} \}
\nonumber \\
&
+ c_{EE} a^2 \{\gamma_4 D_4,
\boldsymbol{\alpha} \cdot \boldsymbol{E}\}
\bigg]
\psi(x)
\nonumber \\
= \int d^4 x ~ \bar{\psi}(x)\bigg[
&m_0 +  \gamma_4D_{4} 
+\zeta \boldsymbol{ \gamma }\cdot \boldsymbol{D}
-\frac{1}{2}aD_4^2 
-\frac{1}{2} r_s \zeta a \boldsymbol{D}^2
-\frac{1}{2} c_B \zeta   i 
 a\boldsymbol{\Sigma} \cdot \boldsymbol{B}
\nonumber \\
&
-\frac{1}{2} c_E \zeta 
a \boldsymbol{\alpha} \cdot \boldsymbol{E} 
+\frac{1}{6} \gamma_4 a^2 D_4^3
+  \big(c_1  +\frac{1}{6}\zeta \big)
\sum_i \gamma_i a^2 D_i^3
+ c_2 a^2 \{  \boldsymbol{\gamma} \cdot \boldsymbol{D}, 
\boldsymbol{D}^2 \}
\nonumber \\
&
+ c_3 a^2 \{ \boldsymbol{\gamma} \cdot \boldsymbol{D} ,
i \boldsymbol{\Sigma} \cdot \boldsymbol{B} \}
+ c_{EE} a^2 \{\gamma_4 D_4,
\boldsymbol{\alpha} \cdot \boldsymbol{E}\}
\bigg]
\psi(x).
\label{eq:symanzik-third}
\end{align}

If the action is to be improved through $\mathcal{O}(a^2)$, the action
in Eq.~\eqref{eq:symanzik-third} should be equivalent to the Dirac
action through $\mathcal{O}(a^2)$,
\begin{align}
\bar{\psi}(x) 
\bar{\mathcal{R}}
\Big[m_q + \boldsymbol{\gamma \cdot D} + \gamma_4 D_4 \Big]
\mathcal{R}\psi(x) = \text{R.H.S of \eqref{eq:symanzik-third}}\,,
\label{eq:sym-matching}
\end{align}
where the transformations $\mathcal{R}$ and $\bar{\mathcal{R}}$ should
be in terms of $m_0a$, $\boldsymbol{\gamma \cdot D}$, and $\gamma_4
D_4$.
To match the action through $\mathcal{O}(a^2)$, they are
\begin{align}
\mathcal{R} &= \bigg[ 1 +\frac{1}{4}m_0a 
- \frac{1}{4}r_s \zeta a \boldsymbol{\gamma \cdot D} 
- \frac{1}{4} a \gamma_4 D_4
- \frac{7}{96}(am_0)^2 
- \frac{1}{48}am_0 (a\gamma_4D_4)
\nonumber \\
&+\Big(\frac{1}{48}+\frac{3r_s \zeta}{16} - \frac{r_s^2 \zeta^2}{16} \Big)
(am_0) a\boldsymbol{\gamma \cdot D}
+\Big(-\frac{1}{48} -\frac{r_s \zeta}{8} + \frac{r_s^2 \zeta^2}{32} \Big)
(a\boldsymbol{\gamma\cdot D})^2
\nonumber \\
&+ \frac{5}{96}(a\gamma_4 D_4)^2 
-\frac{r_s^2 \zeta^2}{32} 
a \gamma_4 D_4
a\boldsymbol{\gamma \cdot D}
+\Big( \frac{1}{48} + \frac{r_s \zeta}{16} -\frac{r_s^2 \zeta^2}{32}
\Big)
a\boldsymbol{\gamma \cdot D} 
a \gamma_4 D_4 \bigg]\,,
\label{eq:R-trans-second}
\\
\bar{\mathcal{R}} &= 
\bigg[ 1 
+\frac{1}{4}m_0a 
- \frac{1}{4}r_s \zeta a \boldsymbol{\gamma \cdot D} - \frac{1}{4} a \gamma_4 D_4
- \frac{7}{96}(am_0)^2 - \frac{1}{48}am_0 (a\gamma_4D_4)
\nonumber \\
&
+\Big(\frac{1}{48}+\frac{3r_s \zeta}{16} - \frac{r_s^2 \zeta^2}{16} \Big)
(am_0) a\boldsymbol{\gamma \cdot D}
+\Big(-\frac{1}{48} -\frac{r_s \zeta}{8} + \frac{r_s^2 \zeta^2}{32} \Big)
(a\boldsymbol{\gamma\cdot D})^2
\nonumber \\
&
+ \frac{5}{96}(a\gamma_4 D_4)^2 
-\frac{r_s^2 \zeta^2}{32} a\boldsymbol{\gamma \cdot D}
a \gamma_4 D_4
+\Big( \frac{1}{48} + \frac{r_s \zeta}{16} -\frac{r_s^2 \zeta^2}{32}
\Big)a \gamma_4 D_4 a\boldsymbol{\gamma \cdot D} \bigg]\,,
\label{eq:R-bar-trans-second}
\end{align}
where the coefficients of Eq.~\eqref{eq:R-trans-second} and
Eq.~\eqref{eq:R-bar-trans-second} are fixed by
Eq.~\eqref{eq:sym-matching}.
For example, the $-\frac{1}{4}a\gamma_4 D_4$ term in
Eq.~\eqref{eq:R-trans-second} and Eq.~\eqref{eq:R-bar-trans-second} is
tuned to fix the coefficient of $aD_4^2$ in
Eq.~\eqref{eq:symanzik-third} to be $-\frac{1}{2}$.
Not only determining Eq.~\eqref{eq:R-trans-second} and
Eq.~\eqref{eq:R-bar-trans-second}, Eq.~\eqref{eq:sym-matching} gives
constraint equations on the action parameters ($\zeta$, $c_B$, $c_E$,
$\cdots$) at the tree level.
For example, if one compares the mass term on both sides of
Eq.~\eqref{eq:sym-matching}, it gives the relation between the
physical quark mass and the bare mass
\begin{align}
m_0 = m_q \Big( 1+\frac{1}{2}m_0a - \frac{1}{12} m_0^2 a^2 \Big),
\end{align}
which gives
\begin{align}
m_q = m_0 - \frac{1}{2}m_0^2 a + \frac{1}{3} m_0^3 a^2.
\label{eq:mass-Sym}
\end{align}
Through second order in $a$, the R.H.S. of Eq.~\eqref{eq:mass-Sym}
is equivalent to the rest mass $m_1 = \text{Log}(1+m_0a)/a$.
Thus, Eq.~\eqref{eq:mass-Sym} is equivalent to identifying the rest
mass with the physical quark mass.
Likewise, if one compares the coefficients of $a\boldsymbol{\gamma
  \cdot D}$ on both sides of Eq.~\eqref{eq:sym-matching}, one obtains
the constraint equation
\begin{align}
1 + \Big( \frac{1}{2} - \frac{1}{2}r_s \zeta  \Big)m_0 a
+ \Big( -\frac{1}{24} + \frac{1}{2}r_s \zeta 
- \frac{1}{8}r_s^2 \zeta^2 \Big) m_0^2 a^2  = \zeta,
\end{align}
which gives
\begin{align}
\zeta = 1 + \frac{1}{2}(1-r_s) m_0a + \frac{1}{24}( -1 +6r_s +3 r_s^2 )m_0^2 a^2
+\mathcal{O}(m_0a)^3,
\label{eq:zeta-2nd}
\end{align}
which is identical to Eq.~(4.11) of Ref.~\cite{ElKhadra:1996mp}.
As mentioned in Ref.~\cite{ElKhadra:1996mp}, the above $\zeta$ value
is determined by the condition $m_1 = m_2$.

Now, if we insert Eq.~\eqref{eq:mass-Sym} and Eq.~\eqref{eq:zeta-2nd}
into the L.H.S. of Eq.~\eqref{eq:sym-matching}, we obtain
\begin{align}
&\bar{R}\Big[m_q + \boldsymbol{\gamma \cdot D} + \gamma_4 D_4 \Big]R
= m_0
+\zeta \boldsymbol{\gamma \cdot D} 
+\gamma_4 D_4
-\frac{1}{2} r_s \zeta a 
\big(\boldsymbol{\gamma \cdot D} \big)^2
\nonumber \\
&
 -\frac{1}{2} a 
\big(\gamma_4 D_4 \big)^2 
+ a\boldsymbol{\alpha \cdot E}
\bigg(- \frac{1}{4} (1+r_s ) + \Big( -\frac{1}{24} + \frac{1}{8}r_s \Big) 
m_0a\bigg)
\nonumber \\
&+ \frac{1}{6}a^2\big(\gamma_4 D_4 \big)^3
+a^2 \big( \boldsymbol{\gamma \cdot D} \big)^3
\Big( -\frac{1}{24} - \frac{r_s}{4}+ \frac{r_s^2}{8} \Big)
\nonumber \\
& + \{ \gamma_4 D_4, \boldsymbol{\alpha \cdot E} \} 
\Big( \frac{5}{96} + \frac{1}{16}r_s \zeta - \frac{1}{32} r_s^2 \zeta^2 \Big),
\end{align}
which determines 
\begin{align}
c_B &= r_s, 
\label{eq:cb-matching}
\\
c_E &= \frac{1}{2}(1+r_s)+\frac{1}{12}\big(-2-3r_s + 3r_s^2\big)m_0 a
+ \mathcal{O}(m_0a)^2, 
\label{eq:ce-matching}
\\
c_1 & = -\frac{1}{6} + \mathcal{O}(m_0a),\\
c_2 &= c_3 = \frac{1}{48}\big(-1-6r_s +3 r_s^2 \big) 
+\mathcal{O}(m_0a),\\
c_{EE} & = \frac{1}{96}\big(5 + 6r_s -3r_s^2 \big)+\mathcal{O}(m_0a).
\label{eq:cee-matching}
\end{align}

The (tree-level) matching of the action through $\mathcal{O}(a^2)$ is
done by specifying the action parameters according to
Eq.~\eqref{eq:mass-Sym}, Eq.~\eqref{eq:zeta-2nd}, and
Eqs.~\eqref{eq:cb-matching}-\eqref{eq:cee-matching}.
If one defines $q(x) = \mathcal{R} \psi(x)$, then the Lagrangian of
$q(x)$ corresponds to the Dirac Lagrangian.

One can identify $\mathcal{R}$ as the transformation required for the
(tree-level) current improvement.
Here we can eliminate terms with the time derivative by using the
equation of motion for the R.H.S. of Eq.~\eqref{eq:sym-matching},
\begin{align}
(a \boldsymbol{\gamma \cdot D}) (a\gamma_4 D_4) \psi(x) 
&=\Big(-(m_0a)(a \boldsymbol{\gamma \cdot D})  
- \zeta (a \boldsymbol{\gamma \cdot D})^2 \Big)\psi(x) ,
\\
(a\gamma_4 D_4)( a \boldsymbol{\gamma \cdot D}) \psi(x) 
&= \Big(a^2 \boldsymbol{\alpha \cdot E} 
+(m_0a)(a \boldsymbol{\gamma \cdot D}) 
+ \zeta (a \boldsymbol{\gamma \cdot D})^2 \Big)\psi(x) ,
\\
(a\gamma_4 D_4 )^2 \psi(x) 
&= \Big(m_0^2 a^2 - \zeta^2 (a\boldsymbol{\gamma \cdot D})^2 
-a^2 \zeta \boldsymbol{\alpha \cdot E}\Big) \psi(x) ,
\\
a\gamma_4 D_4 \psi(x) = 
&\bigg(-m_0a -\zeta a \boldsymbol{\gamma \cdot D}+ \frac{1}{2}r_s \zeta a^2\boldsymbol{D}^2
+\frac{1}{2} c_B \zeta   i 
 a\boldsymbol{\Sigma} \cdot \boldsymbol{B} 
+\frac{1}{2} c_E \zeta   
 a\boldsymbol{\alpha} \cdot \boldsymbol{E} 
\nonumber \\
&
+\frac{1}{2}\Big(m_0^2 a^2 - \zeta^2 (a\boldsymbol{\gamma \cdot D})^2 
-a^2 \zeta \boldsymbol{\alpha \cdot E}\Big)\bigg) \psi(x) .
\end{align}
Then,
%If we factor out prefactor $(1+m_0a)^{1/2}= 1+\frac{1}{2}m_0a +\cdots$ 
%as Eq.~\eqref{eq:improved-current-lambda-third}
%
\begin{align}
\mathcal{R} = 
\Big[1+\frac{1}{2}m_0a -\frac{1}{8}(m_0a)^2\Big]
\Big[ 1&+ \Big( \frac{1}{4}(1-r_s) 
+ \frac{1}{48}\big(1+3r_s^2 \big)m_0 a \Big) a\boldsymbol{\gamma \cdot D}
+ \frac{1}{32}\big( 1- 10r_s + r_s^2 \big) \big(a\boldsymbol{\gamma \cdot D}\big)^2
\nonumber \\
&+ \frac{1}{96}\big(1- 6r_s -3r_s^2 \big) a^2\boldsymbol{\alpha \cdot E}
\Big] + \mathcal{O}\big((m_0a)^3\big),
\label{eq:R-trans-second-a}
\end{align}
which gives the leading behaviors of $d_1$, $d_2$, $d_B$, and $d_E$ as
\begin{align}
d_1 &= \frac{1}{4}(1-r_s) + \frac{1}{48}\big(1 + 3r_s^2 \big)m_0a 
+ \mathcal{O}\big((m_0a)^2\big) 
\label{eq:d1-matching}
\\
d_2 &= d_B  =\frac{1}{32}\big(1-10r_s + r_s^2 \big) + \mathcal{O}(m_0a) \\
d_E &= \frac{1}{48}\big(1-6r_s - 3r_s^2 \big) + \mathcal{O}(m_0a).
\label{eq:de-matching}
\end{align}

\end{widetext}

%-----------
% reference
%-----------
\bibliography{ref} %%% ref.bib file

\end{document}